\documentclass[10pt,final,journal]{IEEEtran}
\usepackage[pdftex]{graphicx}
\usepackage[dvipsnames]{xcolor}
\usepackage{algorithm,algorithmic}
\usepackage{cite}
\usepackage{array}
\usepackage{url}
\usepackage{amsthm}
\usepackage{amsmath}
\usepackage{amssymb}
\usepackage{mathtools}
\usepackage{multicol}
\usepackage{tabularx}
\usepackage{enumitem}
\usepackage{booktabs}
\usepackage{leftidx}
\usepackage{subfig}
\usepackage{tikz}
\DeclareGraphicsExtensions{.pdf,.jpeg,.png}
\theoremstyle{definition}

\newtheorem{theorem}{Theorem}

\newtheorem{corollary}{Corollary}
\newtheorem{lemma}{Lemma}

\DeclareMathOperator*{\argmax}{arg\,max}
\DeclareMathOperator*{\argmin}{arg\,min}
\begin{document}
\title{An Improved Random Matrix Prediction Model for Manoeuvring Extended Targets}
\author{Nathan~J.~Bartlett,
	Chris~Renton,
	Adrian~G.~Wills
	\thanks{All authors are within the Faculty of Engineering and Built Environment,
		The University of Newcastle, Australia.}
	\thanks{Corresponding author e-mail: \protect\url{nathan.bartlett@uon.edu.au}}}
\markboth{}%
{Bartlett \MakeLowercase{\textit{et al.}}}
\maketitle

\begin{abstract}
	This paper proposes an improved prediction update for extended target tracking with the random matrix model.
	A key innovation is to employ a generalised non-central inverse Wishart distribution to model the state transition density of the target extent; resulting in a prediction update that accounts for kinematic state dependent transformations.
	Moreover, the proposed prediction update offers an additional tuning parameter c.f. previous works, requires only a single Kullback-Leibler divergence minimisation, and improves overall target tracking performance when compared to state-of-the-art alternatives.
\end{abstract}
\IEEEpeerreviewmaketitle


\section{Introduction}\label{sec:Introduction}

\IEEEPARstart{T}{arget} tracking is an important practical problem that has received significant research attention since the pioneering works of~\cite{Bar1987a} 
Due to an increase in sensor resolution capabilities, it is now commonplace for multiple measurements to be generated for each target per time-step; e.g., modern radar can produce several range measurements for a single target.
One approach to handle the generation of multiple measurements is to employ extended-target models, where, in addition to estimating the kinematic state variables, the spatial extent of the target is estimated from available data.
In recognition of the utility of extended targets, several models have been proposed over the past decade, see e.g.,~\cite{Gilholm2005a,Boers2006d,Angelova2013a,Mahler2009,Baum2009e,Lundquist2011a,Hirscher2016a,Yang2019,Koch2008a}.

One of the most popular extended-target models, referred to as the \emph{random matrix model}, was first proposed in the works of Koch~\cite{Koch2008a}; and defines the extended target state as the combination of a kinematic state vector and an extent matrix.
The extent matrix is assumed to be symmetric positive definite, which in turn enables for the target extent to be represented by an ellipsoid~\cite{Feldmann2011a}.
The kinematic state vector is modelled as a Gaussian distributed random variable, whilst the extent matrix is modelled as an inverse Wishart distributed random variable.
Following a Bayesian methodology, the prediction density of each random variable is assumed to belong to the same distribution class of its respective posterior, enabling for the parameters of each density to be updated recursively; consisting of a prediction and a correction stage.

In the early work of Koch and Feldmann~\emph{et al.}~\cite{Koch2008a,Feldmann2011a},~the prediction update of the extent matrix was based upon a simple heuristic that artificially increased the covariance whilst preserving the expected value.
Koch additionally proposed the use of a Wishart state transition density~\cite{Koch2008a}, which was later generalised within~\cite{Lan2016,Granstrom2014d} to handle temporal evolutions of the target extent.
However, even for evolutions independent of the kinematic state, a second-order moment matching technique or Kullback-Leibler divergence minimisation is required to approximate the resulting Generalised Beta Type-II prediction density as an inverse Wishart distribution; instigating information loss in regard to the target extent~\cite{Bartlett2020}.
To resolve this issue, Bartlett~\emph{et al.}~\cite{Bartlett2020} proposed the use of a non-central inverse Wishart state transition density, which was shown to produce an inverse Wishart prediction distribution directly and significantly improve target extent estimation.  

\emph{Contributions:} The key contribution of this paper is the generalisation of the non-central inverse Wishart state transition density to account for kinematic state dependent evolutions of the target extent.
This generalisation is motivated by the works of~\cite{Granstrom2014d}, whom, to the best of the authors' knowledge, were the first to propose a prediction update to handle kinematic state dependent evolutions of the target extent.
The main difference here is that by building upon~\cite{Bartlett2020}, our proposed prediction update requires only a \emph{single} Kullback-Leibler divergence minimisation, and offers an additional tuning parameter to model uncertainties in target shape more effectively.

The remainder of the paper is organised as follows.
In Section~\ref{sec:randomMatrix}, we provide an overview of the random matrix model and the prediction updates of~\cite{Koch2008a,Feldmann2011a,Lan2016,Granstrom2014d,Bartlett2020}.
The problem formulation is presented in Section~\ref{sec:ProblemForumlation}, and the proposed prediction update is given in Section~\ref{sec:Prediction}.
Simulated results comparing the proposed prediction update against state-of-the-art alternatives are presented in Section~\ref{sec:Simulations}.
Concluding remarks are given in Section~\ref{sec:conclusions}. 
Notation and distributions are summarised in Table~\ref{table:notation}.


\section{The Random Matrix Model}\label{sec:randomMatrix}

The random matrix model can be accredited to the works of Koch and Feldmann~\textit{et al.}~\cite{Koch2008a,Feldmann2011a}. 
The model has been used in an wide array of extended-target tracking applications over the last decade, including aircraft tracking with ground radar, pedestrian tracking with laser range sensors, and surface vessel tracking with marine X-band radar; see e.g.,~\cite{Granstrom2012d,Granstrom2015a,Granstrom2016c,Granstrom2017f,Vivone2015a,Vivone2019,Schuster2015a,Beard2016b}.

The random matrix model defines the extended target state $\xi_{k} \triangleq (\mathbf{x}_{k},X_{k})$ at time $t_{k}$ as the combination of a kinematic state vector $\mathbf{x}_{k} \in \mathbb{R}^{n_{x}}$ and an extent matrix $X_{k} \in \mathbb{S}^{d}_{++}$.
The kinematic state vector consists of states related to the motion of the 
target centre, whilst the extent matrix models the target extent as a $d$-dimensional ellipsoid~\cite{Bartlett2020}.
Thus, the region of space $\mathcal{Y} \subseteq \mathbb{R}^{d}$ occupied by the extended target at time $t_{k}$ can be described by the following set:
\begin{equation}
	\mathcal{Y} \triangleq \{ \mathbf{y} \in \mathbb{R}^{d} :(\mathbf{y} - H_{k}\mathbf{x}_{k})^{T}X^{-1}_{k}(\mathbf{y} - H_{k}\mathbf{x}_{k}) \leq 1 \},
\end{equation}
where $H_{k}$ is a  matrix that transforms the kinematic state vector to a target position.
In the conditional random matrix model, the kinematic state can only consist of the target position and time derivatives such as velocity and acceleration~\cite{Koch2008a}.
In the factorised random matrix model\textemdash which shall be utilised in this paper\textemdash the kinematic state can also include non-linear kinematics such as turn-rate and heading~\cite{Feldmann2011a}.
The measurements $\mathbf{z}^{r}_{k} \in \mathbf{Z}_{k}$ at time $t_{k}$ are assumed to be multivariate Gaussian distributed with covariance related to the extent matrix 
\begin{equation}\label{eq:originalLikelihood}
	p(\mathbf{Z}_{k}|\xi_{k}) = \prod^{m_{k}}_{r=1} 
	\mathcal{N}(\mathbf{z}^{r}_{k}|H_{k}\mathbf{x}_{k}, \lambda X_{k} + R_{k}),
\end{equation}
where $R_{k}$ is the sensor noise covariance, and $\lambda$ is the scaling factor to the spread contribution of the target extent~\cite{Feldmann2011a}.

In situations where the sensor noise covariance $R_{k}$ has negligible impact upon the spatial distribution of the measurements\footnote{
	$R_{k} \approx 0_{d}$ to prove conjugacy~\cite{Koch2008a}.
	In scenarios where this approximation is invalid, additional heuristics are required to obtain a closed-form correction; see~\cite{Feldmann2011a} for further details.},
Koch showed that an inverse Wishart probability density function, or pdf, is a conjugate prior to~(\ref{eq:originalLikelihood}).
Hence, the factorised random matrix model defines the posterior of the extended target state $\xi_{k}$ as the product of a Gaussian kinematic state density and an inverse Wishart extent matrix density~\cite{Feldmann2011a}.
More specifically, the posterior distribution is approximated as
\begin{subequations} \label{eq:fullposterior}
	\begin{align}
		p(\xi_{k}|\mathbf{Z}^{k}) &\approx p(\mathbf{x}_{k}|,\mathbf{Z}^{k})p(X_{k}|\mathbf{Z}^{k}), \label{eq:posterior}
		\intertext{where $\mathbf{Z}^{k} = \{\mathbf{Z}_{1},\ldots,\mathbf{Z}_{k}\}$ is the set of all measurement sets up to and including time $t_{k}$, and}
		p(\mathbf{x}_{k}|\mathbf{Z}^{k}) &=  \mathcal{N}(\mathbf{x}_{k}|\mathbf{m}_{k|k},P_{k|k}), \label{eq:posteriorx} \\
		p(X_{k}|\mathbf{Z}^{k}) &= \mathcal{IW}_{d}(X_{k}|\nu_{k|k},V_{k|k}).\label{eq:posteriorX}
	\end{align}
\end{subequations}

In Bayesian estimation, it is often desired for the prediction density to belong to the same distribution class as the posterior.
By satisfying this condition, the parameters of the chosen class of distribution can be updated in Bayesian recursion rather than the entire distribution itself.
Therefore, to obtain a computationally efficient filter, we want to ensure that the prediction density is of the same functional form as the posterior~(\ref{eq:fullposterior}); i.e., we want 
\begin{subequations} \label{eq:fullprediction}
	\begin{align}
		p(\xi_{k+1}|\mathbf{Z}^{k}) &\approx p(\mathbf{x}_{k+1}|\mathbf{Z}^{k})p(X_{k+1}|\mathbf{Z}^{k}), \label{eq:predicted}
		\intertext{where}
		p(\mathbf{x}_{k+1}|\mathbf{Z}^{k}) 
		&=\mathcal{N}(\mathbf{x}_{k+1}|\mathbf{m}_{k+1|k},P_{k+1|k}), \label{eq:predictedx} \\
		p(X_{k+1}|\mathbf{Z}^{k}) &=\mathcal{IW}_{d}(X_{k+1}|\nu_{k+1|k},V_{k+1|k}). \label{eq:predictedX}
	\end{align}
\end{subequations}
Assuming this condition is satisfied, the prediction update corresponds to obtaining the parameters $\{\mathbf{m}_{k+1|k}$, $P_{k+1|k}\}$, and $\{\nu_{k+1|k}, V_{k+1|k}\}$ of the predicted Gaussian kinematic state vector density~(\ref{eq:predictedx}) and the predicted inverse Wishart extent matrix density~(\ref{eq:predictedX}) respectively.

\begin{table}[!t]
	\caption{Notation}
	\label{table:notation}
	\begin{tabularx}{\columnwidth}{X}
		\toprule
		\begin{itemize}[wide = 0em]
			\item $I_{n}$ is a $n \times n$ identity matrix, and $0_{m}$ is a $m \times m$ zero matrix.
			\item $\mathbb{R}^{n}$ is the set of real column vectors of length $n$, $\mathbb{R}^{m \times n}$ is the set of real $m \times n$ matrices, $\mathbb{S}^{n}_{++}$ is the set of symmetric positive definite $n \times n$ matrices,  $\mathbb{S}^{n}_{+}$ is the set of symmetric positive semi-definite $n \times n$ matrices, $\mathbb{N}$ is the set of natural numbers, $\mathbb{SL}(n,\mathbb{R})$ is the special linear group of $n \times n$ matrices with determinant equal to one, and $\mathbb{O}^{m \times n}$ is the set of $m \times n$ semi-orthogonal  matrices with $m \leq n$.
			\item $\mathcal{N}(\mathbf{x}|\mathbf{m},P)$ denotes a multivariate Gaussian pdf defined over the vector $\mathbf{x} \in \mathbb{R}^{n}$, with expectation $\mathbf{m} \in \mathbb{R}^{n}$ and covariance matrix $P \in \mathbb{S}^{n}_{++}$,
			\begin{equation*}
				\mathcal{N}(\mathbf{x}|\mathbf{m},P) = \frac{\text{exp}(-\frac{1}{2}(\mathbf{x}-\mathbf{m})^{T}P^{-1}(\mathbf{x}-\mathbf{m}))}{(2\pi)^{\frac{n}{2}}|P|^{\frac{1}{2}}},
			\end{equation*}
			where $|\cdot|$ denotes the matrix determinant.
			\item $\mathcal{IW}_{d}(X|\nu,V)$ denotes an inverse Wishart pdf defined over the matrix $X \in \mathbb{S}^{d}_{++}$ with scalar degrees of freedom $\nu > 2d$ and parameter matrix $V \in \mathbb{S}^{d}_{++}$~\cite[Definition 3.4.1]{Gupta2000},
			\begin{equation*}
				\mathcal{IW}_{d}(X|\nu,V) = \frac{\text{etr}(-\frac{1}{2}VX^{-1})|V|^{\frac{\nu-d-1}{2}}}{2^{\frac{d(\nu-d-1)}{2}}\Gamma_{d}(\frac{\nu-d-1}{2})|X|^{\frac{\nu}{2}}},
			\end{equation*}
			where  $\text{etr}(\cdot) = \exp(\text{Tr}(\cdot))$ represents the exponential of the matrix trace, and $\Gamma_{d}(\cdot)$ is the multivariate Gamma function which can be expressed as a product of ordinary gamma functions $\Gamma(\cdot)$~\cite[Theorem 1.4.1]{Gupta2000}. 
			The expectation of $X$ is $V/(\nu - 2d -2)$~\cite[Theorem 3.4.3]{Gupta2000}.
			\item $\mathcal{W}_{d}(X|w,W)$ denotes a Wishart pdf defined over the matrix $X \in \mathbb{S}^{d}_{++}$ with scalar degrees of freedom $w > d-1$ and parameter matrix $W \in \mathbb{S}^{d}_{++}$ \cite[Definition 3.2.1]{Gupta2000},
			\begin{equation*}
				\mathcal{W}_{d}(X|w,W) = \frac{\text{etr}(-\frac{1}{2}XW^{-1})|X|^{\frac{w-d-1}{2}}}{2^{\frac{wd}{2}}\Gamma_{d}(\frac{w}{2})|W|^{\frac{w}{2}}}.
			\end{equation*}
			\item $\mathcal{GB}^{II}_{d}(X|a,b,\Omega,\Psi)$ denotes a Generalised Beta Type II pdf defined over the matrix $X \in \mathbb{S}^{d}_{++}$ with scalar parameters $a,b > \frac{d-1}{2}$, and matrices $\Omega \in \mathbb{S}^{d}_{++}$, $\Psi \in \mathbb{S}^{d}_{+}$~\cite[Definition 5.2.4]{Gupta2000},
			\begin{equation*}
				\mathcal{GB}^{II}_{d}(X|a,b,\Omega,\Psi) = 
				\frac{|X-\Psi|^{a-\frac{d+1}{2}}|X + 
					\Omega|^{-(a+b)}}{\beta_{d}(a,b)|\Omega+\Psi|^{-b}},
			\end{equation*}
			where $\beta_{d}(a,b)$ is the multivariate beta function, and can be expressed in terms of the multivariate Gamma function $\Gamma_{d}(\cdot)$~\cite[Theorem 1.4.2]{Gupta2000}.
			\item $\mathcal{IW}^{nc}_{d}(X|\nu,\Sigma,\Sigma\Theta)$ denotes a non-central inverse Wishart pdf defined over the matrix $X \in \mathbb{S}^{d}_{++}$ with scalar degrees of freedom $\nu > 2d$, parameter matrix $\Sigma \in \mathbb{S}^{d}_{++}$, and non-centrality parameter matrix $\Theta \in \mathbb{S}^{d}_{+}$ \cite[Definition 3.5.2]{Gupta2000},
			\begin{equation*}
				\mathcal{IW}^{nc}_{d}(X|\nu,\Sigma, \Sigma\Theta) = 
				\frac{\mathcal{IW}_{d}(X|\nu,\Sigma)\text{etr}(-\tfrac{1}{2}\Sigma\Theta)}
				{\leftidx{_0}{F}{_1}(\frac{\nu-d-1}{2};\frac{1}{4}\Sigma\Theta \Sigma X^{-1})^{-1}},
			\end{equation*}
			where $\leftidx{_0}{F}{_1}(\cdot)$ is the hypergeometric function of matrix argument \cite[Theorem~1.6.4]{Gupta2000}.
			When $\Theta = 0_{d,d}$ the distribution reduces to the inverse Wishart distribution.
		\end{itemize} \\
		\bottomrule
	\end{tabularx}
\end{table}

In the early work of Feldmann~\emph{et al.}~\cite{Feldmann2011a}, an extended Kalman filter prediction is utilised to obtain the parameters of the predicted kinematic state vector density~(\ref{eq:predictedx}).
Moreover, the parameters of the predicted extent matrix density~(\ref{eq:predictedX}) are obtained via a simple heuristic that preserves the expected value whilst artificially increasing the covariance\textemdash resembling an 
exponential forgetting of the extent matrix~\cite{Granstrom2016c}.
That is,
\begin{subequations}\label{eq:heuristics}
	\begin{align}
		\nu_{k+1|k} &= 2d + 4 + e^{-T/\tau}(\nu_{k|k} - 2d - 4), 
		\label{eq:heuristicsnu}\\
		V_{k+1|k} &= \frac{\nu_{k+1|k}-2d-2}{\nu_{k|k}-2d-2}V_{k|k}, 
		\label{eq:heuristicsV}
	\end{align}
\end{subequations}
where $T$ is the prediction time interval and $\tau$ the temporal decay constant.
Note that (\ref{eq:heuristicsnu}) is a modified version of the prediction update utilised in~\cite{Koch2008a} to ensure the expected value and covariance of the extent matrix are always well-defined~\cite{Feldmann2008a}.

Koch also proposed to solve the Chapman-Kolmogorov equation for a Wishart state transition density, and then to approximate the resulting Generalised Beta Type-II prediction density as an inverse Wishart distribution~\cite{Koch2008a}.
This idea was further built upon in~\cite{Lan2016}, where an invertible parameter matrix is introduced to describe transformations of the extent matrix that are independent of the kinematic state vector.
Moreover, a second-order moment matching technique is proposed to perform the inverse Wishart approximation~\cite{Lan2016,Lan2012a}.

Granstr\"om \emph{et al.} further generalised the idea of using a Wishart state transition density, presenting a prediction update that enables the evolution of the extent matrix to be functionally dependent upon the kinematic state vector~\cite{Granstrom2014d}:
\begin{equation}
	p(X_{k+1}|\mathbf{x}_{k},X_{k}) = 
	\mathcal{W}_{d}\Big(X_{k+1}|n_{k+1},\frac{M_{\mathbf{x}_{k}}X_{k}M^{T}_{\mathbf{x}_{k}}}{n_{k+1}}\Big).
	\label{eq:granstromWishart}
\end{equation}
Here, the scalar design parameter $n_{k+1} > d-1$, and the matrix transformation $M_{\mathbf{x}_{k}} \triangleq M(\mathbf{x}_{k})$ such that $M \colon \mathbb{R}^{n_{x}} \rightarrow \mathbb{R}^{d \times d}$ is a nonsingular matrix-valued function of the kinematic state vector~\cite{Granstrom2014d}.
In order to account for kinematic state uncertainty, the use of~(\ref{eq:granstromWishart}) leads to the following integral representation of the predicted extent matrix density:
\begin{multline}\label{eq:GranstromInt}
	p(X_{k+1}|\mathbf{Z}^{k}) = \int \mathcal{N}(\mathbf{x}_{k}|\mathbf{m}_{k|k},P_{k|k})\times \\  \mathcal{GB}^{II}_{d}\!\bigg(\! X_{k+1}|\frac{n_{k+1}}{2},\frac{\nu_{k|k}\!-\!2d\!-\!2}{2},\frac{M_{\mathbf{x}_{k}}V_{k|k}M^{T}_{\mathbf{x}_{k}}}{n_{k+1}},0_{d} \bigg) d\mathbf{x}_{k}.
\end{multline}
Unfortunately, the above integral has no analytical solution \cite{Granstrom2014d}.
Hence, a series of Kullback-Leibler divergence minimisations is employed to approximate~(\ref{eq:GranstromInt}) as an inverse Wishart distribution~\cite{Granstrom2011c}.
To summarise:
\begin{equation}
	p(X_{k+1}|\mathbf{Z}^{k}) \approx \mathcal{IW}_{d}(X_{k+1}|\nu_{k+1|k},V_{k+1|k}),
\end{equation}
where, the scalar degrees of freedom $\nu_{k+1|k}$ and parameter matrix $V_{k+1|k}$ are functions of the expected values
\begin{subequations}\label{eq:GranstromExpectation}
	\begin{align}
		C_{1} &= \int \big(M_{\mathbf{x}_{k}}V_{k|k}M_{\mathbf{x}_{k}}^{T}\big)^{-1} \mathcal{N}(\mathbf{x}_{k}|\mathbf{m}_{k|k},P_{k|k}) d\mathbf{x}_{k}, \label{eq:GranstromExpectationI}\\
		C_{2} &=\int M_{\mathbf{x}_{k}}V_{k|k} M^{T}_{\mathbf{x}_{k}} \mathcal{N}(\mathbf{x}_{k}|\mathbf{m}_{k|k},P_{k|k}) d\mathbf{x}_{k}.
	\end{align}
\end{subequations}
Note that the above definition of $C_{1}$ was first presented in~\cite{Granstrom2019}, and results in a more efficient implementation of the prediction update of~\cite{Granstrom2014d}\footnote{ 
	An intermediary Kullback-Leibler divergence minimisation was originally required to approximate the distribution of $M_{\mathbf{x}_{k}}V_{k|k} M^{T}_{\mathbf{x}_{k}}$ as a Wishart distribution~\cite[Section IV.C]{Granstrom2014d}.
	To do so, $C_{1}$ was defined as the expectation of the logarithmic determinant, and a numerical root finding procedure was performed to obtain the degrees of freedom $s_{k+1}$.}.
See~\cite[Table IV]{Granstrom2019} for further details.

Granstr\"om \emph{et al.} showed that the use of~(\ref{eq:granstromWishart}) offers significant improvement in the tracking of extended targets within unknown turn-rates when compared to~\cite{Koch2008a,Feldmann2011a,Lan2016,Lan2012a}.
Nevertheless, even under the assumption that the time evolution is independent of the kinematic state vector, i.e., $M_{\mathbf{x}_{k}} = M_{k}$, the prediction update requires a Kullback-Leibler divergence minimisation or moment matching technique to approximate the Generalised Beta Type-II density as an inverse Wishart distribution~\cite{Granstrom2014d,Granstrom2019}. 
In order to remove the need for such density approximations, Bartlett~\emph{et al.}~\cite{Bartlett2020} proposed the following non-central inverse Wishart state transition density: 
\begin{equation}\label{eq:transitionIWnc}
	\begin{split}
		p(X_{k+1}|&X_{k},\mathbf{Z}^{k}) = \\ &\mathcal{IW}^{nc}_{d}(X_{k+1}|v_{k+1},\Sigma_{k+1},\Sigma_{k+1}\Theta_{k+1}(X_{k})),
	\end{split}
\end{equation}
where the degrees of freedom $v_{k+1}$, parameter matrix $\Sigma_{k+1}$, and non-centrality matrix $\Theta_{k+1}(X_{k})$ are defined as follows:
\begin{subequations}\label{eq:transitionparams}
	\begin{align}
		v_{k+1} &\in (2d,\nu_{k|k}] \cap \mathbb{N},\label{eq:transitionDof}\\ 
		\Sigma_{k+1} &= M_{k+1}Q^{-1}_{k+1}M^{T}_{k+1}, \label{eq:transitionSigma}\\
		\Theta_{k+1}(X_{k}) &= M^{-T}_{k+1}X^{-1}_{k}M^{-1}_{k+1}.
	\end{align}
\end{subequations}
Here, the nonsingular $d\times d$ transition matrix $M_{k+1}$ is used to model transformations independent of the kinematic state vector, and $Q_{k+1} \in \mathbb{S}^{d}_{++}$ is used to model uncertainties in the extent matrix evolution; offering an additional $d(d\!+\!1)/2$ tunable parameters than previous works~\cite{Bartlett2020}. 
Moreover, the use of~(\ref{eq:transitionIWnc}) guarantees the prediction density is of the desired inverse Wishart form~(\ref{eq:predictedX}) with the following scalar degrees of freedom and parameter matrix:
\begin{subequations}
	\begin{align}
		\nu_{k+1|k} &= v_{k+1}, \\
		V_{k+1|k} &= M_{k+1}V_{k|k}(I_{d} + Q_{k}V_{k|k})^{-1}M^{T}_{k+1}.
	\end{align}
\end{subequations}

The main contribution of this paper is to propose a generalisation of~(\ref{eq:transitionIWnc}) that allows for kinematic state dependent evolutions of the extent matrix; e.g., rotations and scaling.
In doing so, the proposed prediction update does not suffer from the same levels of information loss as~\cite{Granstrom2014d,Granstrom2019}, requires only a \emph{single} Kullback-Leibler divergence minimisation, and offers an additional tuning parameter to model uncertainties in target shape more effectively. 


\section{Problem Formulation}\label{sec:ProblemForumlation}

In Bayesian filtering, the prediction step consists of solving the following integral
\begin{equation}\label{eq:Chapman}
	p(\xi_{k+1}|\mathbf{Z}^{k}) = \int p(\xi_{k+1}|\xi_{k},\mathbf{Z}^{k})p(\xi_{k}|\mathbf{Z}^{k}) d\xi_{k}.
\end{equation}
In order to obtain a closed-form solution to this integral, it is assumed in~\cite{Bartlett2020} that the evolution of the extent matrix is independent of the kinematic state vector. 
Albeit true for non-manoeuvring behaviours, this assumption is often violated during constant or variable turn manoeuvres\textemdash in which the target extent rotates as a function of the turn-rate~\cite{Granstrom2014d}. 
Therefore, in order to improve the tracking performance of manoeuvring targets, we will now account for such dependency.

Inspired by~\cite{Granstrom2014d}, the non-Markov state transition density $p(\xi_{k+1}|\xi_{k},\mathbf{Z}^{k})$ is expanded as follows\footnote{
	Following~\cite{Bartlett2020}, the state transition density is non-Markov, and thus retains the dependency upon the measurement set $\mathbf{Z}^{k}$.
	This action enables for greater flexibility in the selection of state transition parameters to model target shape uncertainties; see~\cite[Section IV.D]{Bartlett2020} for further details.}:
\begin{subequations}
	\begin{align}
		\begin{split}
			p(\xi_{k+1}|\xi_{k},\mathbf{Z}^{k}) &= p(\mathbf{x}_{k+1}|X_{k+1},\mathbf{x}_{k},\mathbf{Z}^{k}) \times \\ &\qquad \qquad \qquad p(X_{k+1}|\mathbf{x}_{k},X_{k},\mathbf{Z}^{k}),
		\end{split} \\
		&\approx p(\mathbf{x}_{k+1}|\mathbf{x}_{k})p(X_{k+1}|\mathbf{x}_{k},X_{k},\mathbf{Z}^{k}). \label{eq:StateTransitionApprox}
	\end{align}
\end{subequations}
The evolution of the kinematic state vector is assumed to be independent of the extent matrix.
Hence, phenomena dictated by the target extent are deemed negligible; for example, wind resistance~\cite{Granstrom2016c}.

Given posterior~(\ref{eq:posterior}) and non-Markov state transition density~(\ref{eq:StateTransitionApprox}), equation~(\ref{eq:Chapman}) becomes
\begin{multline}\label{eq:Chapman2}
	p(\xi_{k+1}|\mathbf{Z}^{k}) = \int p(\mathbf{x}_{k+1}|\mathbf{x}_{k}) \int p(X_{k+1}|\mathbf{x}_{k},X_{k}) \\ \times  p(X_{k}|\mathbf{Z}^{k})p(\mathbf{x}_{k}|\mathbf{Z}^{k}) dX_{k} d\mathbf{x}_{k}.
\end{multline}

As previously stated, it is often desired in Bayesian filtering for the prediction density to belong to the same distribution class as the posterior\textemdash enabling for a finite set of statistics to be updated in Bayesian recursion rather than the entire distribution~\cite{Mahler2007b}.
Therefore, given~(\ref{eq:posterior}), we require that the resulting prediction density of~(\ref{eq:Chapman2}) possesses the same functional form as~(\ref{eq:predicted}).
Unfortunately, this condition cannot be proven to hold in general.
Thus, as in the pioneering works of Granstr\"om \emph{et al.}~\cite{Granstrom2014d}, we approximate~(\ref{eq:Chapman2}) to be the product of two independent equations:
one for the kinematic state vector, and another for the extent matrix:
\begin{subequations}\label{eq:Chapman3}
	\begin{align}
		p(\mathbf{x}_{k+1}|\mathbf{Z}^{k}) &= \int p(\mathbf{x}_{k+1}|\mathbf{x}_{k}) p(\mathbf{x}_{k}|\mathbf{Z}^{k}) d\mathbf{x}_{k}, \label{eq:Chapmanx}\\
		\begin{split}
			p(X_{k+1}|\mathbf{Z}^{k}) &= \iint p(X_{k+1}|\mathbf{x}_{k},X_{k},\mathbf{Z}^{k})p(X_{k}|\mathbf{Z}^{k}) \\ & \qquad\qquad\qquad\qquad \times p(\mathbf{x}_{k}|\mathbf{Z}^{k}) dX_{k} d\mathbf{x}_{k}.
		\end{split} \label{eq:ChapmanX}
	\end{align}
\end{subequations}

We must now determine an appropriate state transition density for both the kinematic state vector and extent matrix such that the resulting prediction densities of~(\ref{eq:Chapmanx}) and~(\ref{eq:ChapmanX}) are multivariate Gaussian and inverse Wishart distributions respectively.
Following~\cite{Granstrom2014d,Vivone2016a}, the state transition density of the kinematic state vector is given by,
\begin{equation}\label{eq:KinematicTransition}
	p(\mathbf{x}_{k+1}|\mathbf{x}_{k}) = \mathcal{N}(\mathbf{x}_{k+1}|f_{k+1}(\mathbf{x}_{k}),D_{k+1}),
\end{equation}
where $f_{k+1} \colon \mathbb{R}^{n_{x}} \rightarrow \mathbb{R}^{n_{x}}$ is the nonlinear state transition function, and $D_{k+1} \in \mathbb{S}^{n_{x}}_{++}$ is the dynamic noise covariance matrix~\cite{Li2000a,Li2003a}. 
Substituting the state transition density~(\ref{eq:KinematicTransition}) and posterior~(\ref{eq:posteriorx}) into~(\ref{eq:Chapmanx}), the extended Kalman filter is then used to approximate the resulting prediction density as a multivariate Gaussian distribution~\cite{Schmidt1966a,Smith1962a}. 
That is,
\begin{subequations}
	\begin{align}
		p(\mathbf{x}_{k+1}|\mathbf{Z}^{k}) &\approx \mathcal{N}(\mathbf{x}_{k+1}|\mathbf{m}_{k+1|k},P_{k+1|k}),
		\intertext{where the mean and covariance are given by}
		\mathbf{m}_{k+1|k} &= f_{k+1}(\mathbf{m}_{k|k}),\\
		P_{k+1|k} &= F_{k+1}P_{k|k}F^{T}_{k+1} + D_{k+1}, \\
		F_{k+1} &= \nabla_{\mathbf{x}} f_{k+1}(\mathbf{x}) \Big\rvert_{\mathbf{x} = \mathbf{m}_{k|k}}.
	\end{align}
\end{subequations}

The problem considered in this work is to derive an analogous closed-form prediction update for the extent matrix.
More specifically, to determine a suitable non-Markov state transition density $p(X_{k+1}|\mathbf{x}_{k},X_{k},\mathbf{Z}^{k})$ that can be used in~(\ref{eq:ChapmanX}) to obtain the desired inverse Wishart distribution~(\ref{eq:predictedX}) with minimal approximations.
In addition, the chosen state transition density must offer a high degree of modelling flexibility, be physically interpretable, and encapsulate a wide subset of possible kinematic state dependencies.


\section{Prediction Update}\label{sec:Prediction}

In this section, we introduce the state transition density and present the new prediction update of the extent matrix.
We show that via a single Kullback-Leibler divergence minimisation, the resulting prediction density of~(\ref{eq:ChapmanX}) is an inverse Wishart distribution; given the state transition density is a non-central inverse Wishart distribution.
All supporting lemmata and corollaries are given in the Appendix.


\subsection{The Extent Matrix State Transition Density}\label{sec:Density}

Motivated by~\cite{Granstrom2014d,Bartlett2020}, we define the non-Markov state transition density of the extent matrix as the following non-central inverse Wishart distribution: 
\begin{multline}\label{eq:TransitionIWnc}
	p(X_{k+1}|\mathbf{x}_{k},X_{k},\mathbf{Z}^{k}) =  \\\mathcal{IW}^{nc}_{d}(X_{k+1}|v_{k+1},\Sigma_{k+1}(\mathbf{x}_{k}),\Sigma_{k+1}(\mathbf{x}_{k})\Theta(\mathbf{x}_{k},X_{k})),
\end{multline}
where the scalar degrees of freedom $v_{k+1}$, parameter matrix $\Sigma_{k+1}(\mathbf{x}_{k})$, and non-centrality matrix $\Theta_{k+1}(\mathbf{x}_{k},X_{k})$ are defined as follows:
\begin{subequations}\label{eq:Transitionparams}
	\begin{align}
		v_{k+1} &\in (2d,\nu_{k|k}] \cap \mathbb{N},\label{eq:TransitionDof}\\ 
		\Sigma_{k+1}(\mathbf{x}_{k}) &= M_{\mathbf{x}_{k}}Q^{-1}_{k+1}M^{T}_{\mathbf{x}_{k}}, \label{eq:TransitionSigma}\\
		\Theta_{k+1}(\mathbf{x}_{k},X_{k}) &= M^{-T}_{\mathbf{x}_{k}}X^{-1}_{k}M^{-1}_{\mathbf{x}_{k}}.
	\end{align}
\end{subequations}
The tuneable parameters of the non-Markov state transition density are thereby the degrees of freedom $v_{k+1}$, the symmetric positive definite noise matrix $Q_{k+1} \in \mathbb{S}^{d}_{++}$, and the matrix transformation $M_{\mathbf{x}_{k}} \triangleq M(\mathbf{x}_{k})$; where $M \colon \mathbb{R}^{n_{x}} \rightarrow \mathbb{R}^{d \times d}$ is a nonsingular matrix-valued function of the kinematic state.
To provide adequate meaning to these parameters in regard to extended target tracking, we shall now briefly discuss the underlying state transition model.


\subsection{The Extent Matrix State Transition Model}\label{sec:Model}

The proposed state transition density is a generalisation of \cite{Bartlett2020}.
From Lemma~\ref{lem:transModel}, the state transition model governing~(\ref{eq:TransitionIWnc}) and describing the evolution of the extent matrix from time $t_{k}$ to $t_{k+1}$ is  
\begin{subequations}\label{eq:TransModel}
	\begin{align}
		X^{-\frac{1}{2}}_{k+1} &= M^{-T}_{\mathbf{x}_{k}}\Big(X^{-\frac{1}{2}}_{k} + n_{k+1}^{\frac{1}{2}}W^{\frac{1}{2}}_{k+1}\Big), \label{eq:ProcessModel} \\
		W_{k+1} &\sim \mathcal{W}_{d}\Big(W_{k+1}|n_{k+1},\frac{Q_{k+1}}{n_{k+1}}\Big), \label{eq:Processnoise}
	\end{align}
\end{subequations}
where $n_{k+1} = v_{k+1}\!-\!d\!-\!1$.
Although model~(\ref{eq:TransModel}) seems rather complex, its physical interpretation can be separated into two simple components: the injection of Wishart distributed process noise $W_{k+1}$, and the extent evolution described by the transition matrix $M_{\mathbf{x}_{k}}$~\cite{Bartlett2020}.
We shall now discuss each of these components to highlight the properties of each state transition parameter.

In accordance with~\cite[Theorem 3.3.15]{Gupta2000}, the expectation and variance of the process noise $W_{k+1}$ is given by: 
\begin{subequations}\label{eq:ProcessMoments}
	\begin{align}
		\mathbb{E}[W_{k+1}] &= Q_{k+1}, \label{eq:ProcessExpect}\\
		\text{Var}(W_{k+1}) &= \frac{Q_{k+1}Q_{k+1}}{v_{k+1}\!-\!d\!-\!1} + \frac{\text{tr}(Q_{k+1})Q_{k+1}}{v_{k+1}\!-\!d\!-\!1}. \label{eq:ProcessVariance}
	\end{align}
\end{subequations}
By~(\ref{eq:ProcessMoments}), the primary parameter that governs the size, shape, and expected value of the process noise is the noise matrix $Q_{k+1}$~\cite{Bartlett2020}.
Therefore, contrary to~\cite{Granstrom2014d,Granstrom2019}, our model offers an additional $d(d\!+\!1)/2$ tunable parameters to describe the effects of process noise on the extent matrix evolution. 
For example, varying levels of process noise can be applied to each principle axis of the target extent; which can aid in the tracking of group targets such as truck convoys~\cite{Bartlett2020}.

Furthermore, as $Q_{k+1}$ approaches $0_{d,d}$, the expected value and variance of the process noise also approach $0_{d,d}$~(\ref{eq:ProcessMoments}).
This implies that, although non-linearly, smaller elements of $Q_{k+1}$ results in a more deterministic evolution of the target extent~\cite{Bartlett2020}.
Hence, the noise matrix $Q_{k+1}$ is analogous to the dynamic noise covariance matrix $D_{k+1}$~(\ref{eq:KinematicTransition}); which models the uncertainty in the evolution of the kinematic state vector $\mathbf{x}_{k}$ from time $t_{k}$ to $t_{k+1}$.
In Section~\ref{sec:ParameterSelection}, we shall provide methodologies for tuning $v_{k+1}$ and $Q_{k+1}$ for extended target tracking purposes.

Once the process noise $W_{k+1}$ has been injected, the state transition model then performs the evolution described by $M_{\mathbf{x}_{k}}$.
To highlight the effects of $M_{\mathbf{x}_{k}}$, consider the ideal scenario in which the process noise is zero.
Then,~(\ref{eq:ProcessModel}) is equivalent to the deterministic model
\begin{equation}\label{eq:DeterministicModel}
	X_{k+1} = M_{\mathbf{x}_{k}}X_{k}M^{T}_{\mathbf{x}_{k}}.
\end{equation}
Thus, as in the works of~\cite{Granstrom2014d}, the main motivation for $M_{\mathbf{x}_{k}}$ is to model rotations of the target extent; however, in general, the function $M_{\mathbf{x}_{k}}$ can be selected as any arbitrary transformation\textemdash provided the output is a non-singular $d\times d$ matrix~\cite{Bartlett2020}. 
For example, in group target tracking the group extent may grow or shrink over time, corresponding to $M_{\mathbf{x}_{k}}$ to be a scale matrix~\cite{Granstrom2016c}. 
The matrix transformation $M_{\mathbf{x}_{k}}$ is thereby analogous to the kinematic state transition function $f_{k+1}$~(\ref{eq:KinematicTransition}); which models the evolution of the kinematic state vector $\mathbf{x}_{k}$ from time $t_{k}$ to $t_{k+1}$.
Within our work, we shall consider $M_{\mathbf{x}_{k}}$ to be of the following form:
\begin{equation}\label{eq:Mk}
	M_{\mathbf{x}_{k}} = \begin{bmatrix} \cos(T\omega_{k}) & -\sin(T\omega_{k}) \\ \sin(T\omega_{k}) & \ \  \cos(T\omega_{k}) \end{bmatrix},
\end{equation}
where $T$ is the time interval, $\mathbf{x}_{k} = [x_{k},y_{k},\dot{x}_{k},\dot{y}_{k},\omega_{k}]^{T}$, and $\omega_{k}$ is the turn-rate of the extended target.
Note that~(\ref{eq:Mk}) is commonly utilised to model target extent rotations; see e.g.,~\cite{Granstrom2014d,Granstrom2019,Granstrom2015a,Vivone2016a}. 


\subsection{The Generalised Extent Matrix Prediction Update}\label{sec:ExtentPrediction}

Substituting the posterior~(\ref{eq:fullposterior}) and the state transition density (\ref{eq:TransitionIWnc}) into~(\ref{eq:ChapmanX}), the prediction update of the extent matrix is equivalent to
\begin{multline}
	p(X_{k+1}|\mathbf{Z}^{k}) = \\
	\iint \mathcal{IW}^{nc}_{d}(X_{k+1}|v_{k+1},\Sigma_{k+1}(\mathbf{x}_{k}),\Sigma_{k+1}(\mathbf{x}_{k})\Theta(\mathbf{x}_{k},X_{k}))\\ 
	\times \mathcal{N}(\mathbf{x}_{k}|\mathbf{m}_{k|k},P_{k|k}) \mathcal{IW}_{d}(X_{k}|\nu_{k|k},V_{k|k}) dX_{k} d\mathbf{x}_{k},
\end{multline}
which, given~\cite[Theorem 1]{Bartlett2020}, yields the intermediate integral
\begin{subequations}
	\begin{multline}
		p(X_{k+1}|\mathbf{Z}^{k}) =  \int \mathcal{N}(\mathbf{x}_{k}|\mathbf{m}_{k|k},P_{k|k}) \\ \times \mathcal{IW}_{d}(X_{k}|v_{k+1},M_{\mathbf{x}_{k}}\bar{V}_{k+1}M^{T}_{\mathbf{x}_{k}}) d\mathbf{x}_{k}, \label{eq:PredictionInterim}
	\end{multline}
	where the intermediate parameter matrix $\bar{V}_{k+1} \in \mathbb{S}^{d}_{++}$ is
	\begin{equation}
		\bar{V}_{k+1} = V_{k|k}(I_{d} + Q_{k+1}V_{k|k})^{-1}. \label{eq:VInterm}
	\end{equation}
\end{subequations}
Unfortunately, the above integral~(\ref{eq:PredictionInterim}) has no analytical solution.
Hence, motivated by~\cite{Granstrom2014d}, we resort to approximating the resulting prediction density with an inverse Wishart distribution through Kullback-Leibler divergence minimisation. 

Kullback-Leibler divergence is considered the optimal difference measure when approximating distributions in a maximum likelihood sense~\cite{Williams2003a,Runnalls2007a,Hershey2007a}.
For the two distributions $p(X)$ and $q(X)$, the Kullback-Leibler divergence is defined as follows~\cite{Kullback195b}:
\begin{equation}\label{kldivergence}
	\text{KL}\big(p(X) || q(X)\big) = \int p(X)\log \bigg(\frac{p(X)}{q(X)}\bigg) dX.
\end{equation}
By defining $p(X)$ as~(\ref{eq:PredictionInterim}) and $q(X)$ by~(\ref{eq:predictedX}), Lemma~\ref{lem:KLD} proves that the best and unique global approximation of~(\ref{eq:PredictionInterim}) that minimises~(\ref{kldivergence}) is given by
\begin{subequations}
	\begin{align}
		p(X_{k+1}|\mathbf{Z}^{k}) &\approx \mathcal{IW}_{d}(X_{k+1}|\nu_{k+1|k},V_{k+1|k}), \label{eq:PredictionIW}
		\intertext{where the parameter matrix $V_{k+1|k}$ is equal to}
		V_{k+1|k} &= \bigg(\frac{\nu_{k+1|k}\!-\!d\!-\!1}{v_{k+1}\!-\!d\!-\!1}\bigg)C^{-1}_{1}, \label{eq:PredictionV}
	\end{align}
	and the scalar degrees of freedom $\nu_{k+1|k}$ is the solution to
	\begin{multline}\label{eq:nuPred}
		d\ln\bigg(\frac{\nu_{k+1|k}\!-\!d\!-\!1}{v_{k+1}\!-\!d\!-\!1}\bigg) + \sum^{d}_{i=1}\psi_{0}\bigg(\frac{v_{k+1}\!-\!d\!-\!i}{2}\bigg) \\ - \sum^{d}_{i=1}\psi_{0}\bigg(\frac{\nu_{k+1|k}\!-\!d\!-\!i}{2}\bigg) - C_{3} - \ln\big(|C_{1}|\big) = 0. 
	\end{multline}
	Furthermore, $C_{1}$ and $C_{3}$ are the expectations
	\begin{align}
		C_{1} &= \int \big(M_{\mathbf{x}_{k}}\bar{V}_{k+1}M_{\mathbf{x}_{k}}^{T}\big)^{-1} \mathcal{N}(\mathbf{x}_{k}|\mathbf{m}_{k|k},P_{k|k}) d\mathbf{x}_{k}, \label{eq:C1}\\
		C_{3} &= \int \ln\big(|M_{\mathbf{x}_{k}}\bar{V}_{k+1}M_{\mathbf{x}_{k}}^{T}|\big) \mathcal{N}(\mathbf{x}_{k}|\mathbf{m}_{k|k},P_{k|k})   d\mathbf{x}_{k}. \label{eq:C3}
	\end{align}
\end{subequations}
The optimal value of the scalar degrees of freedom $\nu_{k+1|k}$ can be found by applying a numerical root-finding algorithm to~(\ref{eq:nuPred}).
Examples include the Newton-Raphson algorithm and Halley's method~\cite{Ortega1970a}.
Nevertheless, to improve the computational efficiency of the prediction update, we shall utilise the following theorem to obtain a closed-form solution for $\nu_{k+1|k}$.

\begin{theorem}\label{thm:NuPred}
	The expected values of~(\ref{eq:PredictionInterim}) and~(\ref{eq:PredictionIW}) will have the same volume if the scalar degrees of freedom $\nu_{k+1|k}$ is
	\begin{equation}\label{eq:nuClosedForm}
		\nu_{k+1|k} = 2d + 2 + \frac{(d+1)\rho_{k+1}}{(\rho_{k+1}\!+\!d\!+\!1)|C_{1}C_{2}|^{\frac{1}{d}}\!-\!\rho_{k+1}},
	\end{equation}
	where $\rho_{k+1} =v_{k+1}\!-\!2d\!-\!2$, and the expectation
	\begin{equation}\label{eq:C2}
		C_{2} = \int M_{\mathbf{x}_{k}}\bar{V}_{k+1}M_{\mathbf{x}_{k}}^{T} \mathcal{N}(\mathbf{x}_{k}|\mathbf{m}_{k|k},P_{k|k}) d\mathbf{x}_{k}.
	\end{equation}	
	\begin{proof}
		Let $\bar{\mathbb{E}}[X_{k+1}|\mathbf{Z}^{k}]$ and $\mathbb{E}[X_{k+1}|\mathbf{Z}^{k}]$ denote the expected values of~(\ref{eq:PredictionInterim}) and~(\ref{eq:PredictionIW}) respectively.
		By~\cite[Theorem 3.4.3]{Gupta2000},
		\begin{subequations}
			\begin{align}
				\bar{\mathbb{E}}[X_{k+1}|\mathbf{Z}^{k}] &= \frac{C_{2}}{v_{k+1}-2d-2}, \\ 
				\mathbb{E}[X_{k+1}|\mathbf{Z}^{k}] &= \frac{V_{k+1|k}}{\nu_{k+1|k}-2d-2}.
			\end{align}
		\end{subequations}
		The volume of an ellipsoid $X$ is equal to $c_{d}|X|$ where $c_{d}$ is the volume of a $d$-dimensional unit sphere~\cite{Bartlett2020}.
		Matching the volume of the expected values is thereby equivalent to matching the determinants
		\begin{equation}\label{eq:VolumeDets}
			\left\lvert\frac{V_{k+1|k}}{\nu_{k+1|k}-2d-2}\right\rvert = \left\lvert\frac{C_{2}}{v_{k+1}-2d-2}\right\rvert.
		\end{equation}
		Substituting~(\ref{eq:PredictionV}) into~(\ref{eq:VolumeDets}) and re-arranging yields~(\ref{eq:nuClosedForm}).
	\end{proof}
\end{theorem}

As stated previously, the main motivation for $M_{\mathbf{x}_{k}}$ is to model rotations of the target extent.
Thus, to compare the optimal~(\ref{eq:nuPred}) and closed-form~(\ref{eq:nuClosedForm}) solutions of $\nu_{k+1|k}$, we define $M_{\mathbf{x}_{k}}$ by~(\ref{eq:Mk}).
Furthermore, motivated by~\cite[Corollary 2]{Granstrom2014d}, we approximate the expectations $C_{1}$, $C_{2}$, and $C_{3}$ with third-order Taylor series expansions.
That is, given $\mathbb{V}_{\mathbf{x}_{k}} \triangleq M_{\mathbf{x}_{k}}\bar{V}_{k+1}M^{T}_{\mathbf{x}_{k}}$:
\begin{subequations}\label{eq:GranstromTaylor}
	\begin{align}
		C_{1} &\!\approx\! \bigg(\!\mathbb{V}^{-\!1}_{\mathbf{x}_{k}} +  \sum^{n_{x}}_{i=1} \sum^{n_{x}}_{j=1} \frac{\partial^{2} \mathbb{V}^{-\!1}_{\mathbf{x}_{k}}}{\partial \mathbf{x}^{i}_{k}\partial \mathbf{x}^{j}_{k}} P^{ij}_{k|k} \bigg)\bigg\rvert_{\mathbf{x}_{k}=\mathbf{m}_{k|k}}, \\
		C_{2} &\!\approx\! \bigg(\!\mathbb{V}_{\mathbf{x}_{k}}  + \sum^{n_{x}}_{i=1} \sum^{n_{x}}_{j=1}  \frac{\partial^{2} \mathbb{V}_{\mathbf{x}_{k}}}{\partial \mathbf{x}^{i}_{k}\partial \mathbf{x}^{j}_{k}} P^{ij}_{k|k} \bigg)\bigg\rvert_{\mathbf{x}_{k}=\mathbf{m}_{k|k}}, \\
		C_{3} \!&\approx\! \bigg(\!\ln\!\big(|\mathbb{V}_{{\mathbf{x}}_{k}}|\big)+ \sum^{n_{x}}_{i=1} \sum^{n_{x}}_{j=1}  \frac{\partial^{2} \ln\big(|\mathbb{V}_{\mathbf{x}_{k}}|\big)}{\partial \mathbf{x}^{i}_{k}\partial \mathbf{x}^{j}_{k}} P^{ij}_{k|k} \bigg)\bigg\rvert_{\mathbf{x}_{k}=\mathbf{m}_{k|k}},
	\end{align}
\end{subequations}
where $\mathbf{x}^{i}_{k}$ denotes the $i^{\text{th}}$ element of $\mathbf{x}_{k}$, and $P^{ij}_{k|k}$ denotes the $(i,j)^{\text{th}}$ element of $P_{k|k}$.

\begin{figure*}[!t]
	\centering
	\subfloat[]{\includegraphics[width=0.85\columnwidth]{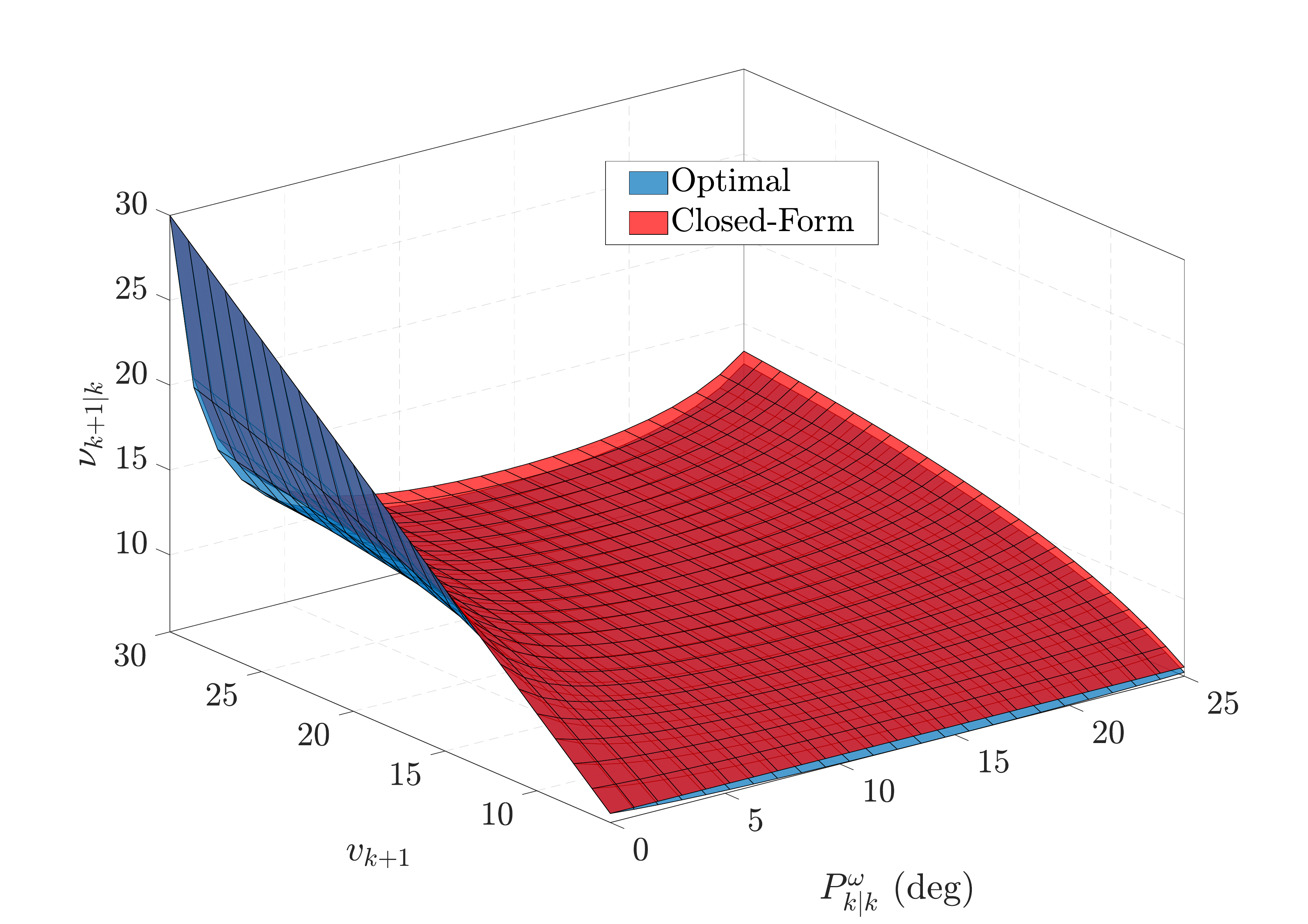}%
		\label{fig:NuOptimalApprox}}\hfill
	\centering
	\subfloat[]{\includegraphics[width=0.85\columnwidth]{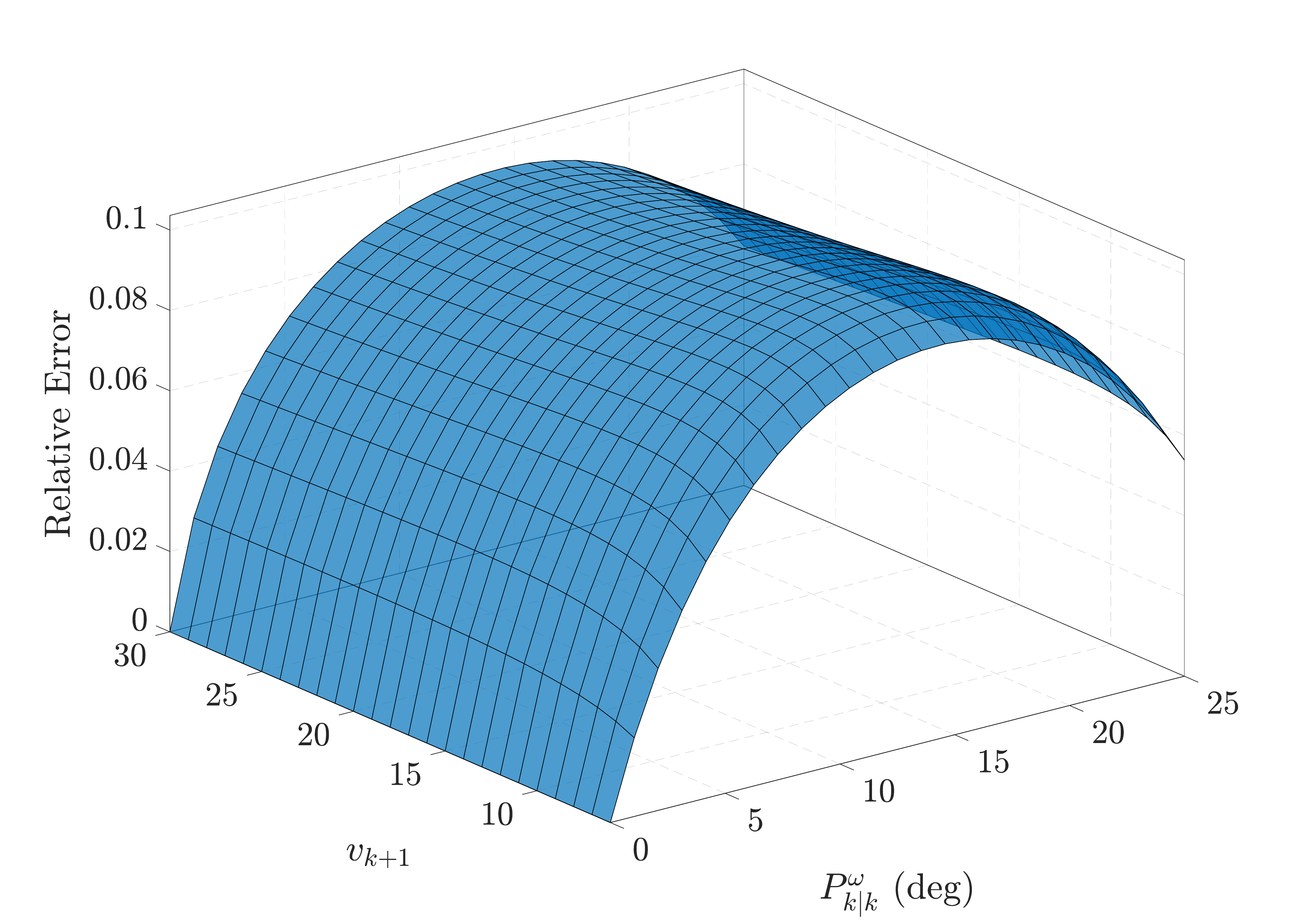}%
		\label{fig:NuError}}
	\caption{(a) The optimal (blue) and closed-form (red) solutions of $\nu_{k+1|k}$ for differing levels of turn-rate variance $P^{\omega}_{k|k}$ and state transition degrees of freedom $v_{k+1}$.
		(b) The relative error between the optimal and closed-form solutions.}
	\label{fig:NuPredict}
\end{figure*}

Figure~\ref{fig:NuPredict} shows the optimal and closed-form solutions of $\nu_{k+1|k}$ for differing levels of turn-rate variance $P^{\omega}_{k|k}$\footnote{
	By~(\ref{eq:Mk}), the matrix transformation $M_{\mathbf{x}_{k}}$ is a function of the turn-rate $\omega_{k}$ and prediction time interval $T$.
	Therefore, no covariance element other than turn-rate variance will impact the expectations $C_{1}$, $C_{2}$, and $C_{3}$.
} 
and state transition degrees of freedom $v_{k+1}$.
The intermediate parameter matrix $\bar{V}_{k+1} = \text{diag}([10^2,5^2])$, and the expected value of the turn-rate $\omega_{k|k} = 10^{\circ}/T$.
From Figure~\ref{fig:NuError}, it is observed that the maximum relative error between the optimal and closed-form solutions is less than ten percent.
We remark here that relative error experienced between the optimal and closed-form solutions proposed in~\cite{Granstrom2014d} is \emph{typically on the order of one tenth of a degree of freedom}; see~\cite[Corollary 1]{Granstrom2014d}.
We thereby conclude that, as in the works of~\cite{Granstrom2014d}, our closed-form solution is an adequate approximation to the optimal solution.

In addition to the above discussion, the closed-form solution~(\ref{eq:nuClosedForm}) ensures the volume of the expected value of the extent matrix remains constant over the inverse Wishart approximation.
Furthermore, by Jensen's inequality, the expected value of the extent matrix is always well-defined.
That is, the closed-form solution guarantees $\nu_{k+1|k} > 2d + 2$.
We remark here that the same cannot be said for the optimal solution; which only guarantees $\nu_{k+1|k} > 2d$.
The proposed prediction update is presented in Table~\ref{algorithm:Prediction}.
To avoid numerical root-finding and ensure the expected value is well-defined, the update uses Theorem~\ref{thm:NuPred}.

\begin{table}[!t]
	\caption{Proposed Prediction Update}
	\label{algorithm:Prediction}
	\begin{tabularx}{\columnwidth}{X}
		\toprule
		\textbf{Input:} 
		Previous kinematic state estimate, covariance, degrees of freedom, and parameter matrix $\{\mathbf{m}_{k|k},P_{k|k},\nu_{k|k},V_{k|k}\}$.
		Kinematic state transition function and covariance $\{f_{k+1},D_{k+1}\}$.
		Extent matrix state transition parameters and transformation $\{v_{k+1},Q_{k+1},M_{\mathbf{x}_{k}}\}$. \\
		\textbf{Output:} {$\{\mathbf{m}_{k+1|k},P_{k+1|k},\nu_{k+1|k},V_{k+1|k}\}$.}
		{	\begin{subequations}
				\begin{align*}
					\mathbf{m}_{k+1|k} &= f_{k+1}(\mathbf{m}_{k|k}), \\
					P_{k+1|k} &= F_{k+1}P_{k|k}F^{T}_{k+1} + D_{k+1}, \\
					\nu_{k+1|k} &= 2d + 2 + \frac{(d+1)\rho_{k+1}}{(\rho_{k+1}\!+\!d\!+\!1)|C_{1}C_{2}|^{\frac{1}{d}}\!-\!\rho_{k+1}},\\
					V_{k+1|k} &=  \bigg(\frac{\nu_{k+1|k}-d-1}{v_{k+1}-d-1}\bigg)C^{-1}_{1}, \\
					\bar{V}_{k+1} &= V_{k|k}(I_{d}+Q_{k+1}V_{k|k})^{-1}, \\
					\rho_{k+1} &= v_{k+1}-2d-2, \\
					F_{k+1} &= \nabla_{\mathbf{x}} f_{k+1}(\mathbf{x}) \Big\rvert_{\mathbf{x} = \mathbf{m}_{k|k}}, \\
					C_{1} &= \int \big(M_{\mathbf{x}_{k}}\bar{V}_{k+1}M_{\mathbf{x}_{k}}^{T}\big)^{-1} \mathcal{N}(\mathbf{x}_{k}|\mathbf{m}_{k|k},P_{k|k}) d\mathbf{x}_{k}, \\
					C_{2} &= \int M_{\mathbf{x}_{k}}\bar{V}_{k+1}M_{\mathbf{x}_{k}}^{T} \mathcal{N}(\mathbf{x}_{k}|\mathbf{m}_{k|k},P_{k|k}) d\mathbf{x}_{k}.
				\end{align*}
		\end{subequations}}
		\\
		\bottomrule
	\end{tabularx}
\end{table}


\subsection{Parameter Selection}\label{sec:ParameterSelection}

In this section, we discuss methodologies to select the state transition parameters to model practical phenomena.
Before presenting said methodologies however, we first remark that the extent matrix state transition density~(\ref{eq:TransitionIWnc}) is non-Markov.
This enables for the state transition parameters $v_{k+1}$ and $Q_{k+1}$ to be functionally dependent upon the posterior parameters $V_{k|k}$ and $\nu_{k|k}$.

The first setting introduced for the state transition degrees of freedom is
\begin{equation}\label{eq:nuSetting1}
	v_{k+1} = 2d + 2 + (\nu_{k|k}-2d-2)|I_{d} + Q_{k+1}V_{k|k}|^{-\frac{1}{d}}.
\end{equation}

In accordance with~\cite{Bartlett2020}, the above setting ensures that the volume of the expected value of the extent matrix  is preserved for all $Q_{k+1}$.
Nevertheless, the volume is still dependent upon the matrix transformation $M_{\mathbf{x}_{k}}$.
Specifically, by Lemma~\ref{lem:VolumeResults},
\begin{equation}
	\begin{split}\label{eq:VolPreserve1}
		&\text{Vol}\big(\mathbb{E}[X_{k+1}|\mathbf{Z}^{k}]\big) = \\&\text{Vol}\bigg(\!\int\! M_{\mathbf{x}_{k}}\mathbb{E}[X_{k}|\mathbf{Z}^{k}]H  M_{\mathbf{x}_{k}}^{T} \mathcal{N}(\mathbf{x}_{k}|\mathbf{m}_{k|k},P_{k|k}) d\mathbf{x}_{k} \bigg),
	\end{split}
\end{equation}
where $H \triangleq (I_{d}+Q_{k+1}V_{k|k})^{-1}|I_{d}+Q_{k+1}V_{k|k}|^{\frac{1}{d}} \in \mathbb{SL}(d,\mathbb{R})$. 

From~(\ref{eq:VolPreserve1}), it can be seen that the variance of the kinematic state vector influences the volume of the expected value.
As the variance increases, so to will the expected volume.
The setting is thereby suitable for modelling changes in target size that are dependent upon the kinematic state vector.
For example, the spread of a truck convoy may be dependent upon the convoy speed or acceleration\footnote{In general, the distance between each truck will increase as a function of the convoy speed or acceleration.
	Hence, the overall size of the truck convoy must be modified accordingly.}.
Assuming $M_{\mathbf{x}_{k}}$ is an appropriate scale matrix, equation~(\ref{eq:nuSetting1}) ensures that the size of the truck convoy is adequately updated based upon the current estimate of the kinematic state.
Similarly, the shape of the truck convoy can be adequately modified through the noise matrix $Q_{k+1}$ to handle any unforeseen transformations of the group extent.
We refer the reader to~\cite[Section IV-D]{Bartlett2020} for further details on the selection of noise matrix $Q_{k+1}$ to model practical phenomena.

In contrast to the above discussion, the volume of an extended target may be time invariant.
For example, the volume of a surface vessel or ground vehicle will not change over time.
Hence, it would be desirable to ensure the volume of the expected value of the extent matrix is preserved from time $t_{k}$ to $t_{k+1}$.
This gives motivation for the following setting of the state transition degrees of freedom;
\begin{equation}\label{eq:nuSetting2}
	v_{k+1} = 2d + 2 + (\nu_{k|k}-2d-2)|C^{-1}_{2}V_{k|k}|^{-\frac{1}{d}}.
\end{equation}
The above setting ensures that the volume of the expected value of the extent matrix is preserved for all $Q_{k+1}$ and $M_{\mathbf{x}_{k}}$.
That is,
\begin{equation}\label{eq:VolPreserve2}
	\text{Vol}\big(\mathbb{E}[X_{k+1}|\mathbf{Z}^{k}]\big)= \text{Vol}\big(\mathbb{E}[X_{k}|\mathbf{Z}^{k}]\big).
\end{equation}
Given setting~(\ref{eq:nuSetting2}), it is assumed that the kinematic state vector only influences the shape and orientation of the target extent.
All influences in size are neglected.
We remark here that any unforeseen evolutions in target shape can still be adequately modelled by the noise matrix $Q_{k+1}$.
Furthermore, if $M_{\mathbf{x}_{k}}$ is independent of $\mathbf{x}_{k}$ and $|M_{\mathbf{x}_{k}}| = 1$,~(\ref{eq:nuSetting2}) reduces to~(\ref{eq:nuSetting1}).


\section{Simulations}\label{sec:Simulations}

This section presents several simulated results that compare the proposed prediction update to~\cite{Granstrom2014d,Bartlett2020}.
The works of~\cite{Lan2016} have been excluded from this section, as the prediction update of~\cite{Bartlett2020} was shown to outperform~\cite{Lan2016} whenever the turn-rate is unknown to the observer; see~\cite[Section VII]{Bartlett2020} for further discussion.


\subsection{Constant Turn Manoeuvre}\label{sec:ConstantTurnSim}

\begin{figure}[!t]
	\centering
	\includegraphics[width=\columnwidth]{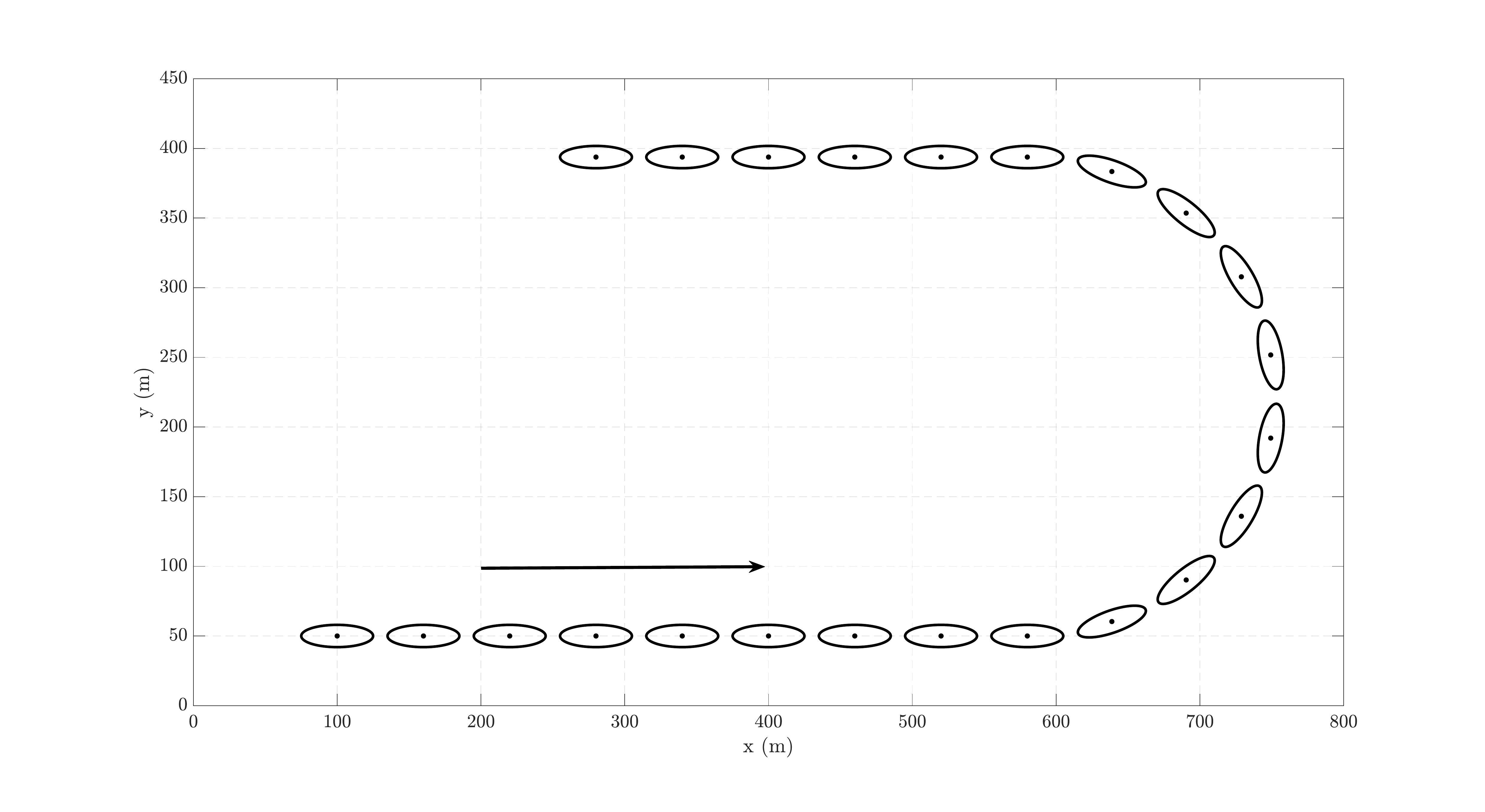}
	\caption{Trajectory of an extended target.
		Shown are, for every second scan $k$, the true target position and the true target extent.}
	\label{fig:Eotpath}
\end{figure}

A single extended target followed the trajectory shown in Figure~\ref{fig:Eotpath}.
The target extent was represented by an ellipsoid with diameters $50$m and $16$m respectively.
The speed of the extended target was assumed constant at $30$m/s.
At time-step $k=18$, the extended target performed a constant turn manoeuvre with a turn-rate of $10^{\circ}$/s.
Scattering measurement centers were uniformly distributed over the target extent $X_{k}$, and the measurement noise was assumed to be zero-mean Gaussian distributed with covariance $R_{k} = \text{diag}([1.5^{2},1.5^{2}])$.
The number of measurements at each time step was Poisson distributed with mean 10, and the scan rate was $T = 1$s.

\begin{figure*}[!t]
	\centering
	\includegraphics[width=0.95\textwidth]{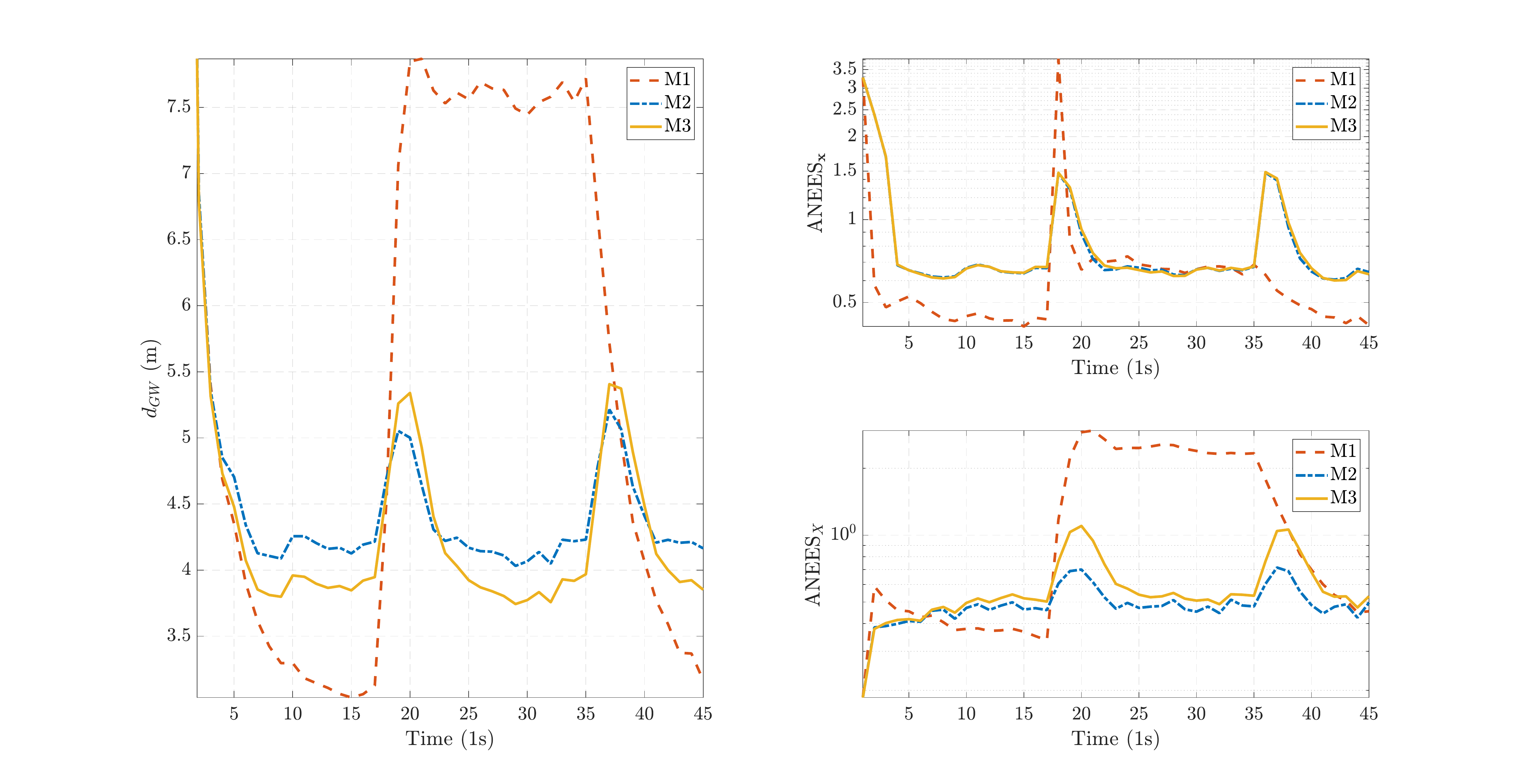}
	\caption{The Gaussian Wasserstein distance (left) and ANEES (right) over 900 Monte Carlo runs.
		The red dashed line is approach M1, the blue dash-dot line is approach M2 and the yellow solid line is approach M3.}
	\label{fig:wassersteinCircle}
\end{figure*}

Herein, the prediction of~\cite{Bartlett2020} is denoted by M1.
A white noise acceleration model was used to update the kinematic state vector; which contained the Cartesian position and velocity of the extended target.
That is, $\mathbf{x}_{k} = [x_{k},y_{k},\dot{x}_{k},\dot{y}_{k}]^{T}$.
We remark here that M1 assumes a conditional random matrix model, therefore, unlike its competitors, the turn-rate of the extended target cannot be included within the kinematic state vector.
The three discrete models of M1 were chosen to adopt the following parameters\footnote{
	Here, $\tilde{q}$ denotes the process noise intensity, and is required when using the discrete-time equivalent white noise acceleration model; see~\cite{Li2003a} for details.
	In accordance with Feldmann \emph{et al.}~\cite{Feldmann2011a}, we define $\tilde{q} \triangleq 0.75T\sigma^{2}_{a}$, where $\sigma_{a}$ denotes the standard deviation of the target acceleration.}:
\begin{enumerate}
	\item $\pi_{1}:$ Low kinematic and extent process noise; $\tilde{q} = 0.001$m$^{2}$/s$^{3}$ and $Q_{k+1} = 0.2V^{-1}_{k|k}$,
	\item $\pi_{2}$: Medium kinematic and extent process noise; $\tilde{q} = 3$m$^{2}$/s$^{3}$ and $Q_{k+1} = 0.33V^{-1}_{k|k}$,
	\item $\pi_{3}$: High kinematic and extent process noise; $\tilde{q} = 6.75$m$^{2}$/s$^{3}$ and $Q_{k+1} = 1.25|V_{k|k}|^{-\frac{1}{d}}I_{d}$.
\end{enumerate}
It was assumed no prior information was known in regard to the turn-rate manoeuvre.
Thus, the transition matrix $M_{k+1}$ was chosen to be~(\ref{eq:Mk}) with $\sigma_{k+1} = 1$ and $\theta = 0^{\circ}$ for each model.

Henceforth, the estimators using the prediction update of~\cite{Granstrom2014d} and the proposed prediction update are denoted by M2 and M3 respectively.
To remove any potential bias, the correction step of Feldmann \emph{et al.}~\cite{Feldmann2011a} was used by both estimators.
Furthermore, a constant turn model was used to update the kinematic state vector of M2 and M3; which contained the Cartesian position, velocity, and turn-rate of the extended target.
That is, $\mathbf{x}_{k} = [x_{k},y_{k},\dot{x}_{k},\dot{y}_{k},\omega_{k}]^{T}$.
The parameters of the two estimators were chosen as follows:
\begin{enumerate}
	\item M2: Medium kinematic and extent process noise; $\sigma_{a} = 2.0$, $\sigma_{\omega} = 0.1^{\circ}$, and $n_{k+1} = 30$,
	\item M3: Medium kinematic and extent process noise; $\sigma_{a} = 2.0$, $\sigma_{\omega} = 0.1^{\circ}$, and $Q_{k+1} = 0.33V^{-1}_{k|k}$,
\end{enumerate}
where $\sigma_{a}$ and $\sigma_{\omega}$ denote the standard deviation of the target acceleration and turn-rate respectively; see~\cite{Li2003a} for details.
Furthermore, for M3, the state transition degrees of freedom $v_{k+1}$ was set according to~(\ref{eq:nuSetting2}).

To evaluate the tracking performance of each estimator, we exploit two credibility measures over $N = 900$ Monte Carlo runs.
The first credibility measure is the Gaussian Wasserstein distance $d_{GW}$, which is the recommended metric for comparing elliptical extended targets~\cite{Granstrom2016c,Yang2016a}.
The square of the Gaussian Wasserstein distance is given by: 
\begin{equation}
\begin{split}
&d^{2}_{GW} \!=\!\frac{1}{N}\sum^{N}_{i=1} \Big(\text{tr}(X_{k} \!+\! \bar{X}_{k|k} \!-\! 2\hat{X}_{k}) \!+\! \left\lVert \mathbf{m}_{k|k} \!-\! \mathbf{x}_{k} \right\rVert^{2}\Big),
\end{split}
\end{equation}
where $\bar{X}_{k|k} = V_{k|k}/(\nu_{k|k}\!-\!2d\!-\!2)$, and $\hat{X}_{k} = (X^{\frac{1}{2}}_{k}\bar{X}_{k|k}X^{\frac{1}{2}}_{k})^{\frac{1}{2}}$.

The second credibility measure is the average normalised estimation error squared (ANEES), which measures how confident an estimator is in its estimation quality~\cite{Feldmann2011a}.
Values greater than one indicate that the estimator is overly confident, whilst values less than one indicate that the estimator is too pessimistic.
The ANEES of the kinematic state vector and the extent matrix are calculated as follows:
\begin{subequations}
	\begin{align}
	\!\!\!\text{ANEES}_{\mathbf{x}} &=  \frac{1}{Nn_{x}}\sum^{N}_{i=1}(\mathbf{m}_{k|k}-\mathbf{x}_{k})^{T}P_{k|k}(\mathbf{m}_{k|k}-\mathbf{x}_{k}), \\
	\!\!\!\text{ANEES}_{X}& = \frac{1}{N}\sum^{N}_{i=1} \frac{\text{Tr}((X_{k|k} - X_{k})^2)}{e_{k|k}},
	\end{align}
\end{subequations}
where $e_{k|k} \triangleq \text{Tr}(\text{Var}(X_{k}|\mathcal{Z}^{k}))$~\cite{Feldmann2011a}.

Figure~\ref{fig:wassersteinCircle} shows the results of each estimator.
Due to the use of a white noise acceleration model, M1 offers the best target tracking performance during the constant velocity motion.
Nevertheless, during the turn-rate manoeuvre, both M2 and M3 offer significant improvement in target tracking performance than M1.
This performance increase can be attributed to two key points.
Firstly, the use of a constant-turn model to update the kinematic state vector, and secondly, the use of the matrix transformation $M_{\mathbf{x}_{k}}$~(\ref{eq:Mk}) to update the orientation of target extent.
Moreover, the ANEES of the target extent is significantly larger for M1 than the other two estimators.
This indicates that M1 experiences greater levels of overconfidence in its estimation quality than M2 and M3; particularly for the target extent.


\subsection{Variable Turn Manoeuvre}

\begin{figure}[!t]
	\centering
	\includegraphics[width=0.95\columnwidth]{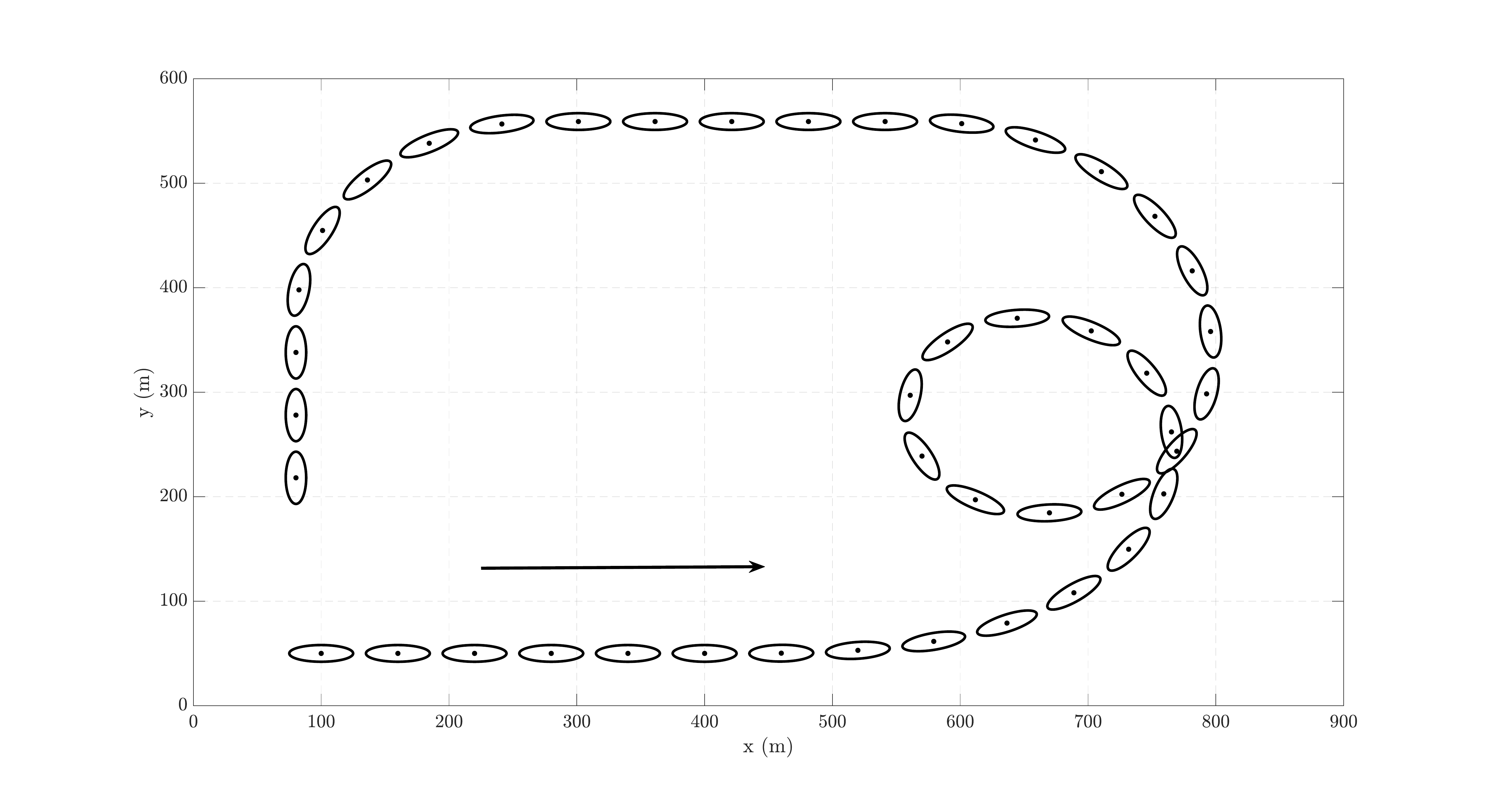}
	\caption{Trajectory of an extended target.
		Shown are, for every second scan $k$, the true target position and the true target extent.}
	\label{fig:Eotpath2}
\end{figure}

In this simulation, the extended target followed the trajectory presented in Figure~\ref{fig:Eotpath2}.
As this trajectory involves long sequences of variable and constant turn-manoeuvres, we have restricted our simulation to compare only the factorised estimators M2 and M3.
The parameters of each estimator were set to the same values as in the previous simulation.

\begin{figure*}[!t]
	\centering
	\includegraphics[width=0.95\textwidth]{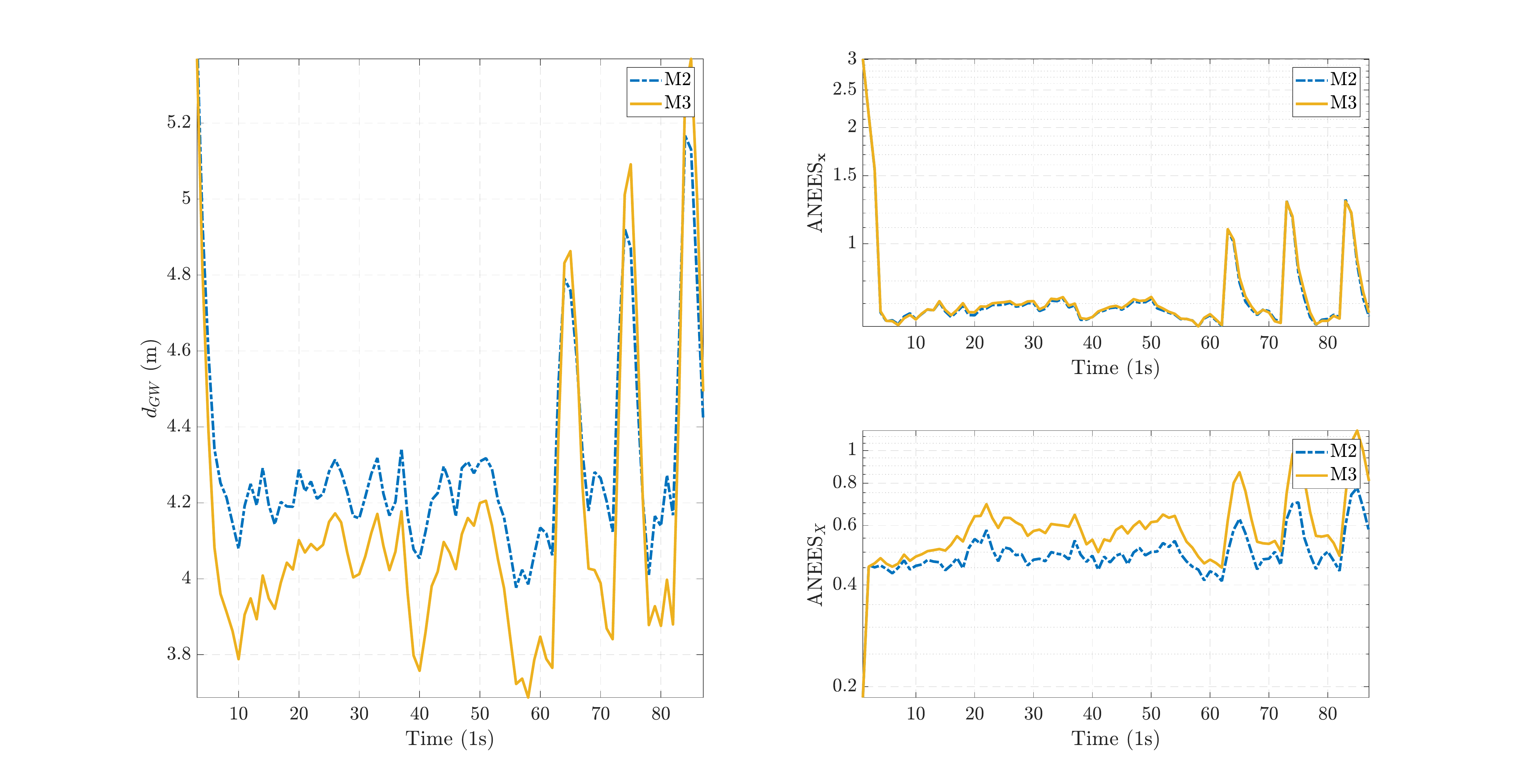}
	\caption{The Gaussian Wasserstein distance (left) and ANEES (right) over 900 Monte Carlo runs.
		The blue dash-dot line is approach M2 and the yellow solid line is approach M3.}
	\label{fig:wassersteinVary}
\end{figure*}

Figure~\ref{fig:wassersteinVary} shows the results of each estimator.
Interestingly, M3 possesses a lower Gaussian Wasserstein distance than M2 for a significant portion of the simulation, with a maximum relative difference of~$\approx 8.7\%$ (left).
This could be a benefit of using a single Kullback-Leibler\ divergence minimisation as opposed to several, or due to preserving the volume of the expected extent matrix via the state transition degrees of freedom $v_{k+1}$.
In regard to the ANEES, both estimators appear to be too over-confident in their estimation quality of the kinematic state vector (top right); particularly in the last sequence of constant velocity and constant turn motions.
For the target extent, the ANEES of both estimators never exceeds a value of one.
We also note that the ANEES of M3 is much larger than M2; with a maximum relative difference of~$\approx 32.7\%$ (bottom right).
This indicates that M3 is less pessimistic than M2, and therefore possesses a better level of confidence in its estimation quality of the target extent.


\section{Concluding Remarks}\label{sec:conclusions}

In this contribution, we generalised the prediction update presented~\cite{Bartlett2020} to enable for kinematic state dependent evolutions of the target extent.
In contrast to~\cite{Granstrom2014d}, the newly proposed update required the use of only a \emph{single} Kullback-Leibler divergence minimisation, and offers an additional tuning parameter to model target shape uncertainties\textemdash the noise matrix $Q_{k+1}$.
To avoid numerical root-finding, we additionally provided a closed-form solution for the predicted degrees of freedom $\nu_{k+1|k}$.
Further, we presented two alternatives on the selection of the state transition degrees of freedom $v_{k+1}$ to model practical phenomena.
Simulated results indicate that the proposed prediction update enables for improved target tracking performance than previous works.

Although not utilised to its full potential, the noise matrix can be utilised to model transformations of the target extent.
Future research should be pushed in this direction to determine other potential structures of $Q_{k+1}$ which can aid in the tracking of extended and group targets.


\appendix 

\begin{lemma}\label{lem:transModel}
	Let the non-Markov state transition density $p(X_{k+1}|\mathbf{x}_{k}, X_{k},\mathbf{Z}^{k})$ be the non-central inverse Wishart distribution described in~(\ref{eq:TransitionIWnc}).
	Then, the state transition model governing the evolution of $X_{k}$ to $X_{k+1}$ is
	\begin{subequations}\label{eq:basismodel}
		\begin{align}
			X^{-\frac{1}{2}}_{k+1} &= M^{-T}_{\mathbf{x}_{k}}\Big(X^{-\frac{1}{2}}_{k} + n_{k+1}^{\frac{1}{2}}W^{\frac{1}{2}}_{k+1}\Big), \\
			W_{k+1} &\sim \mathcal{W}_{d}\Big(W_{k+1}|n_{k+1},\frac{Q_{k+1}}{n_{k+1}}\Big), 
		\end{align}
	\end{subequations}
	where $n_{k+1} \triangleq v_{k+1}-d-1$.
	\begin{proof}
		Let~$X^{-1}_{k} = Y_{k}Y^{T}_{k}$, where $Y_{k}$ is a real $d \times n_{k}$ matrix with dimension $n_{k} \triangleq \nu_{k|k}-d-1$.
		By~\cite[Lemma 2]{Bartlett2020},
		\begin{equation}\label{eq:rootX}
			X^{-1}_{k} = Y_{k}Y^{T}_{k} =( Y_{k}H^{T}_{1})(Y_{k}H^{T}_{1})^{T},
		\end{equation}
		where $H_{1} \in \mathbb{O}^{n_{k+1}\times n_{k}}$ with $n_{k+1} \triangleq v_{k+1}\!-\!d\!-\!1$.	
		Similarly, let $X^{-1}_{k+1} = Y_{k+1}Y^{T}_{k+1}$, where $Y_{k+1}$ is a real $d \times n_{k+1}$ matrix.
		Then, using~\cite[Lemma 3]{Bartlett2020},~(\ref{eq:TransitionIWnc}) is equivalent to
		\begin{equation}\label{eq:transDensityY}
			\begin{split}
				p(Y_{k+1}|&\mathbf{x}_{k},Y_{k},\mathbf{Z}^{k}) = \\ &\mathcal{N}_{d,n_{k+1}}(Y_{k+1}|M^{-T}_{\mathbf{x}_{k}}Y_{k}H^{T}_{1},\Sigma^{-1}_{\mathbf{x}_{k}}\!\otimes\!I_{n_{k+1}}),
			\end{split}
		\end{equation}
		where $\Sigma_{\mathbf{x}_{k}} = M_{\mathbf{x}_{k}}Q^{-1}_{k+1}M^{T}_{\mathbf{x}_{k}}$~(\ref{eq:TransitionSigma}).
		The state transition model governing~(\ref{eq:transDensityY}) is given by:
		\begin{subequations}
			\begin{align}
				Y_{k+1} &= M^{-T}_{\mathbf{x}_{k}}\Big(Y_{k}H^{T}_{1} + n^{\frac{1}{2}}_{k+1}U_{k+1}\Big), \\
				U_{k+1} &\sim \mathcal{N}_{d,n_{k+1}}\Big(U_{k+1}|0_{d,n_{k+1}},\frac{Q_{k+1}}{n_{k+1}} \otimes I_{n_{k+1}}\Big).
			\end{align}
		\end{subequations}
		Define $X^{-\frac{1}{2}}_{k} = Y_{k}H^{T}_{1}$, $X^{-\frac{1}{2}}_{k+1} = Y_{k+1}$, and $W^{\frac{1}{2}}_{k+1} = U_{k+1}$.
		Then, according to~\cite[Theorem 3.2.2]{Gupta2000}, the state transition model governing the evolution of $X_{k}$ to $X_{k+1}$ is given by:
		\begin{subequations}
			\begin{align}
				X^{-\frac{1}{2}}_{k+1} &= M^{-T}_{\mathbf{x}_{k}}\Big(X^{-\frac{1}{2}}_{k} + n_{k+1}^{\frac{1}{2}}W^{\frac{1}{2}}_{k+1}\Big), \\
				W_{k+1} &\sim \mathcal{W}_{d}\Big(W_{k+1}|n_{k+1},\frac{Q_{k+1}}{n_{k+1}}\Big);
			\end{align}
		\end{subequations}
		which is the desired result.
	\end{proof}
\end{lemma}


\begin{lemma}\label{lem:KLD}
	Consider the probability density function
	\begin{equation}\label{eq:PX}
	p(X) = \int \mathcal{IW}_{d}(X|v,M_{\mathbf{x}}\bar{V}M^{T}_{\mathbf{x}}) \mathcal{N}(\mathbf{x}|\mathbf{m},P) d\mathbf{x},
	\end{equation}
	where $M_{\mathbf{x}} \triangleq M(\mathbf{x})$ such that $M\colon \mathbb{R}^{n_{x}} \rightarrow \mathbb{R}^{d\times d}$ is a non-singular matrix-valued function.
	Moreover, let $(\nu^{*},V^{*})$ denote the parameter set that minimises the Kullback-Leibler \ divergence between $p(X)$ and all inverse Wishart distributions $\mathcal{IW}_{d}$; i.e.,
	\begin{equation}
	(\nu^{\ast},V^{\ast}) = \argmin_{(\nu,V)} \ \text{\normalfont KL}\big(p(X) || \mathcal{IW}_{d}(X|\nu,V)\big).
	\end{equation}
	Then, the matrix $V^{\ast}$ is given by
	\begin{equation}\label{eq:Vast}
	V^{\ast} = \bigg(\frac{\nu^{\ast}\!-\!d\!-\!1}{v\!-\!d\!-\!1}\bigg)C^{-1}_{1},
	\end{equation}
	and $\nu^{\ast}$ is the solution to
	\begin{multline}\label{eq:nuast}
	d\ln\bigg(\frac{\nu^{*}\!-\!d\!-\!1}{v\!-\!d\!-\!1}\bigg) + \sum^{d}_{i=1}\psi_{0}\bigg(\frac{v\!-\!d\!-\!i}{2}\bigg) \\- \sum^{d}_{i=1}\psi_{0}\bigg(\frac{\nu^{*}\!-\!d\!-\!i}{2}\bigg) - C_{3} - \ln\big(|C_{1}|\big) = 0, 
	\end{multline}
	where $C_{1}$ and $C_{3}$ are the expectations:
	\begin{subequations}
		\begin{align}
		C_{1} &= \int \big(M_{\mathbf{x}}\bar{V}M^{T}_{\mathbf{x}}\big)^{-1} \mathcal{N}(\mathbf{x}|\mathbf{m},P)   d\mathbf{x},\\
		C_{3} &= \int \ln\big(|M_{\mathbf{x}}\bar{V}M^{T}_{\mathbf{x}}|\big) \mathcal{N}(\mathbf{x}|\mathbf{m},P)  d\mathbf{x}.
		\end{align}
	\end{subequations}
	\begin{proof}
		By equation~(\ref{kldivergence}),
		\begin{subequations}
			\begin{align}
			(\nu^{\ast},V^{\ast}) &= \argmin_{(\nu,V)} \ \text{KL}\big(p(X) || \mathcal{IW}_{d}(X|\nu,V)\big), \\
			&\equiv \argmax_{(\nu,V)} \int p(X)\ln\big(\mathcal{IW}_{d}(X|\nu,V)\big) \ dX, \\
			&\equiv \argmax_{(\nu,V)} f(\nu,V),
			\end{align}
		\end{subequations}
		where the objective function 		
		\begin{multline}\label{eq:ObjectiveFunction}
		f(\nu,V) = \frac{\nu\!-\!d\!-\!1}{2}\Big(\!\ln\big(|V|\big) - d\ln(2)\Big) - \frac{\nu}{2}\mathbb{E}\big[\ln\big(|X|\big)\big] \\ - \frac{1}{2}\text{tr}\big(V\mathbb{E}\big[X^{-1}\big]\big) -\ln\Big(\Gamma_{d}\Big(\frac{\nu\!-\!d\!-\!1}{2}\Big)\Big).
		\end{multline}
		The first-order necessary condition for optimality requires that the gradient of the object function~(\ref{eq:ObjectiveFunction}) will be zero.
		Therefore, $(\nu^{*},V^{*})$ must satisfy
		\begin{equation}
		\nabla f(\nu,V) =  \left. \begin{bmatrix}
		\dfrac{\partial f(\nu,V)}{\partial \text{vec}(V)^{T}} & \dfrac{\partial f(\nu,V)}{\partial \nu}
		\end{bmatrix}^{T} \right\rvert	_{(\nu^{\ast},V^{\ast})} \!\!= 0.
		\end{equation}
		Using the notation of Magnus and Neudecker~\cite{Magnus1985a}, the derivative of the objective function with respect to $\text{vec}(V)$ is
		\begin{subequations}\label{eq:GradientV}
			\begin{align}
			\frac{\partial f(\nu,V)}{\partial \text{vec}(V)} &= \frac{\nu\!-\!d\!-\!1}{2}\frac{\partial \ln\big(|V|\big)}{\partial \text{vec}(V)} - \frac{1}{2}\frac{\partial \text{tr}\big(V\mathbb{E}\big[X^{-1}\big]\big)}{\partial \text{vec}(V)}, \\
			&= \frac{\nu\!-\!d\!-\!1}{2}\text{vec}(V^{-1}) - \frac{1}{2}\text{vec}(\mathbb{E}\big[X^{-1}\big]). \label{eq:GradientV2}
			\end{align}
		\end{subequations}
		Setting~(\ref{eq:GradientV2}) to zero and using Lemma~\ref{lem:ExpectationInverse}:
		\begin{equation}\label{eq:Vast-Expect}
		V^{\ast} = \bigg(\frac{\nu^{\ast}-d-1}{v-d-1}\bigg)C^{-1}_{1},
		\end{equation}
		which is the desired result~(\ref{eq:Vast}).
		Similarly, the derivative of the objective function with respect to $\nu$ is:	
		\begin{multline}
		\frac{\partial f(\nu,V)}{\partial \nu} = \frac{1}{2}\Big(\!\ln\big(|V|\big) - d\ln(2)-\mathbb{E}\big[\ln\big(|X|\big)\big]\Big) \\ -\frac{\partial}{\partial\nu}\ln\Big(\Gamma_{d}\Big(\frac{\nu\!-\!d\!-\!1}{2}\Big)\Big), \label{eq:Gradientnu}
		\end{multline}
		which, through the use of Lemma~\ref{lem:LnDetExpectation} and~\cite[Theorem 1.4.1]{Gupta2000} is equivalent to
		\begin{multline}
		\frac{\partial f(\nu,V)}{\partial \nu} = \frac{1}{2}\Big(\!\ln\big(|V|\big) + \sum^{d}_{i=1}\psi_{0}\bigg(\frac{v\!-\!d\!-\!i}{2}\bigg) \\-\sum^{d}_{i=1}\psi_{0}\Big(\frac{\nu\!-\!d\!-\!i}{2}\Big) - C_{3}\Big). \label{eq:Gradientnu2}
		\end{multline}
		Substitution of~(\ref{eq:Vast-Expect}) into~(\ref{eq:Gradientnu2}) and solving for zero yields~(\ref{eq:nuast}).
		
		To complete the proof, we must now show that the objective function is concave, and that $(\nu^{*},V^{*})$ is the unique global maximum.
		For the objective function to be concave, the hessian matrix, denoted by $\nabla^{2} f(\nu,V)$, must be symmetric negative definite, or conversely $-\nabla^{2} f(\nu,V) \in \mathbb{S}^{(d^{2}+1)}_{++}$, where		
		\begin{equation}\label{eq:Hessian}
		\nabla^{2} f(\nu,V) = \begin{bmatrix}		
		\dfrac{\partial^{2} f(\nu,V)}{\partial \text{vec}(V) \partial\text{vec}(V)^{T}} & \dfrac{\partial^{2} f(\nu,V)}{\partial \text{vec}(V) \partial \nu}  \\[10pt]
		\dfrac{\partial^{2} f(\nu,V)}{\partial \nu \partial \text{vec}(V)^{T}} & \dfrac{\partial^{2} f(\nu,V)}{\partial \nu \partial \nu}  
		\end{bmatrix}.
		\end{equation}	
		Using equations~(\ref{eq:GradientV2}),~(\ref{eq:Gradientnu2}) and~\cite[Section 8.4]{Magnus1999}, the negative of the hessian matrix is equal to
		\begin{equation}
		-\nabla^{2} f(\nu,V) \!=\! \!\begin{bmatrix}
		\dfrac{(V^{-1}\!\otimes\!V^{-1})}{2(\nu\!-\!d\!-\!1)^{-1}}  & -\dfrac{1}{2}\text{vec}(V^{-1}) \\[10pt]
		-\dfrac{1}{2}\text{vec}(V^{-1})^{T} & \dfrac{1}{4}\!\sum\limits_{i=1}^{d}\! \psi_{1}\Big(\dfrac{\nu\!-\!d\!-\!i}{2}\Big)
		\end{bmatrix}\!\!,
		\end{equation}	
		where the Kronecker product is symmetric positive definite for all $\nu > 2d$ and $V \in \mathbb{S}^{d}_{++}$.
		Hence, using the Schur complement\footnote{We refer the reader to~\cite{Zhang2006a} for further details on the Schur complement.}, $-\nabla^{2} f(\nu,V)$ is symmetric positive definite if and only if
		\begin{multline}\label{eq:SchurInequality}
		\sum\limits_{i=1}^{d} \psi_{1}\Big(\frac{\nu\!-\!d\!-\!i}{2}\Big) > \frac{2}{\nu\!-\!d\!-\!1}\\\times \text{vec}(V^{-1})^{T}(V \otimes V)\text{vec}(V^{-1}),
		\end{multline}
		which holds true by Corollary~\ref{cor:TrigammaInequality}.
		Hence, the objective function is concave, and $(\nu^{\ast},V^{\ast})$ is the global minimum of the Kullback-Leibler divergence between $p(X)$ and all $\mathcal{IW}_{d}$ distributions.
		Moreover, this minimum is unique.
	\end{proof}
\end{lemma}


\begin{lemma}\label{lem:ExpectationInverse}
	Let $p(X)$ be the probability density function described in~(\ref{eq:PX}).
	Then,
	\begin{equation}
	\mathbb{E}\big[X^{-1}\big] = (v\!-\!d\!-\!1) C_{1},
	\end{equation}
	where $C_{1}$ denotes the expectation
	\begin{equation}\label{eq:C1a}
	C_{1} = \int \big(M_{\mathbf{x}}\bar{V}M^{T}_{\mathbf{x}}\big)^{-1}\mathcal{N}(\mathbf{x}|\mathbf{m},P) d\mathbf{x}.
	\end{equation}
	\begin{proof}
		Let $X = W^{-1}$. 
		By~\cite[Theorem 3.4.1]{Gupta2000},
		\begin{subequations}
			\begin{align}
			\mathbb{E}\big[W\big] &= \int W p(W) dW,\\
			\begin{split}
			\mathbb{E}\big[W\big] &= \iint W \mathcal{W}_{d}(W|v\!-d\!-1,(M_{\mathbf{x}}\bar{V}M^{T}_{\mathbf{x}})^{-1}) \\& \qquad\qquad\qquad\qquad\quad \ \ \times \mathcal{N}(\mathbf{x}|\mathbf{m},P)  dW d\mathbf{x},
			\end{split} \\
			&= \int (v\!-d\!-1)(M_{\mathbf{x}}\bar{V}M^{T}_{\mathbf{x}})^{-1} \mathcal{N}(\mathbf{x}|\mathbf{m},P)  d\mathbf{x},
			\end{align}
		\end{subequations}
		where the last equality is a consequence of~\cite[Theorem 3.3.15]{Gupta2000}.
		Substitution of~(\ref{eq:C1a}) yields the desired result.
	\end{proof}
\end{lemma}


\begin{lemma}\label{lem:LnDetExpectation}
	Let $p(X)$ be the probability density function described in~(\ref{eq:PX}).
	Then,
	\begin{equation}\label{eq:ExpectationLnDet}
	\mathbb{E}\big[\ln\big(|X|\big)\big] = C_{3} - d\ln(2) - \sum^{d}_{i=1} \psi_{0}\Big(\frac{v\!-\!d\!-\!i}{2}\Big),
	\end{equation}
	where $C_{3}$ denotes the integral
	\begin{equation}\label{eq:C3a}
	C_{3} = \int \ln\big(|M_{\mathbf{x}}\bar{V}M^{T}_{\mathbf{x}}|\big) \mathcal{N}(\mathbf{x}|\mathbf{m},P)  d\mathbf{x}.
	\end{equation}
	\begin{proof}
		By the law of the unconscious statistician~\cite[pg. 274]{Billingsley1995a},
		\begin{subequations}
			\begin{align}
			\mathbb{E}\big[\ln\big(|X|\big)\big] &= \int \ln\big(|X|\big) p(X) dX, \\
			\begin{split}
			 &= \iint \ln\big(|X|\big) \mathcal{IW}_{d}(X|v,M_{\mathbf{x}}\bar{V}M^{T}_{\mathbf{x}}) \\ &\qquad\qquad\qquad\quad \times \mathcal{N}(\mathbf{x}|\mathbf{m},P)  dX d\mathbf{x}.
			\end{split}\label{eq:ExpectationLnDet1}
			\end{align}
		\end{subequations}
		Thus, in order to prove~(\ref{eq:ExpectationLnDet}), we must first find an analytical expression for the following intermediate expectation:
		\begin{equation}
		\mathbb{E}_{\mathcal{IW}}\big[\!\ln\!\big(|X|\big)\big] \triangleq \int\! \ln\!\big(|X|\big) \mathcal{IW}_{d}(X|v,M_{\mathbf{x}}\bar{V}M^{T}_{\mathbf{x}}) dX. \label{eq:interimExpect}
		\end{equation}		
		Using the moment generating function,~(\ref{eq:interimExpect}) is equivalent to the following partial derivative:
		\begin{equation}
		\mathbb{E}_{\mathcal{IW}}\big[\!\ln\!\big(|X|\big)\big] = \frac{\partial}{\partial s}\mathbb{E}_{\mathcal{IW}}\big[|X|^{s}\big] \bigg\rvert_{s = 0}, \label{eq:ExpectationLogDeterminantInverseWishartMGF}
		\end{equation}
		where the expectation		
		\begin{subequations}
			\begin{align}
			\mathbb{E}_{\mathcal{IW}}\big[|X|^{s}\big] &= \int |X|^{s} \mathcal{IW}_{d}(X|v,M_{\mathbf{x}}\bar{V}M^{T}_{\mathbf{x}}) dX, \\
			&= \frac{\Gamma_{d}\Big(\frac{v-2s-d-1}{2}\Big)}{\Gamma_{d}\Big(\frac{v-d-1}{2}\Big)}\frac{|M_{\mathbf{x}}\bar{V}M^{T}_{\mathbf{x}}|^{s}}{2^{ds}}. \label{eq:ExpectationLogDeterminantInverseWishartDetS}
			\end{align}
		\end{subequations}
		Substituting~(\ref{eq:ExpectationLogDeterminantInverseWishartDetS}) into~(\ref{eq:ExpectationLogDeterminantInverseWishartMGF}), and using Lemmata~\ref{lem:MultivariateGammaDerivative} and~\ref{lem:ExpectationDS}:
		\begin{multline}\label{eq:ExpectationIW}
		\!\!\!\!\mathbb{E}_{\mathcal{IW}}\big[\!\ln\!\big(|X|\big)\big] \!= \ln\!\bigg(\frac{|M_{\mathbf{x}}\bar{V}M^{T}_{\mathbf{x}}|}{2^{d}}\bigg) -\! \sum^{d}_{i=1} \psi_{0}\Big(\frac{v\!-\!d\!-\!i}{2}\Big).
		\end{multline}
		Substitution of~(\ref{eq:ExpectationIW}) into~(\ref{eq:ExpectationLnDet1}) yields~(\ref{eq:ExpectationLnDet}).
	\end{proof}
\end{lemma} 


\begin{lemma}\label{lem:VecValue}
	Let $V \in \mathbb{S}^{d}_{++}$ where $d \in \mathbb{N}$.
	Then,
	\begin{equation}\label{eq:vecResult}
	\text{vec}(V^{-1})^{T}(V \otimes V)\text{vec}(V^{-1}) = d.
	\end{equation}
	\begin{proof}	
		From~\cite[Theorem 1.2.22]{Gupta2000}
		\begin{equation}
		\text{vec}(V^{-1})^{T}(V \otimes V)\text{vec}(V^{-1}) = \text{vec}(V^{-1})^{T}\text{vec}(V). \label{firstVec}
		\end{equation}
		\begin{subequations}	
		Moreover, by~\cite[Theorem 1.2.10]{Gupta2000}, there exists an orthogonal matrix $H_{1} \in \mathbb{O}^{d\times d}$ and diagonal matrix $\Lambda = \text{diag}(\lambda_{1},\ldots,\lambda_{d})$ such that
		$V = H_{1}\Lambda H^{T}_{1}$.
		Therefore,
		\begin{align}
			\!\!\text{vec}(V^{-1})^{T}\text{vec}(V) &= \text{vec}(H_{1}\Lambda^{-1} H^{T}_{1})^{T}\text{vec}(H_{1}\Lambda H^{T}_{1}), \\
			&= \text{vec}(\Lambda^{-1})^{T}(I_{d} \otimes I_{d})\text{vec}(\Lambda), \\
			&= \sum^{d}_{i=1} \frac{1}{\lambda_{i}}\lambda_{i} \\&= d. \label{lastVec}
			\end{align}
		\end{subequations}
	Substitution of~(\ref{lastVec}) into~(\ref{firstVec}) yields~(\ref{eq:vecResult}).
	\end{proof}	
\end{lemma}


\begin{lemma}\label{lem:TrigammaInequality}
	Let $\nu > 2d$ where $d \in \mathbb{N}$.
	Then,
	\begin{equation}\label{eq:TrigammaInequality}
	\sum^{d}_{i=1} \psi_{1}\Big(\frac{\nu\!-\!d\!-\!i}{2}\Big) > \frac{2d}{\nu\!-\!d\!-\!1}.
	\end{equation}
	\begin{proof}
		Consider the following upper bound:
		\begin{equation}\label{eq:UpperBound}
		\sum^{d}_{i=1} \frac{2}{\nu\!-\!d\!-\!i} > \frac{2d}{\nu\!-\!d\!-\!1}.
		\end{equation}
		Thus, rather then solving~(\ref{eq:TrigammaInequality}) directly, it suffices to prove
		\begin{equation}\label{eq:SummationInequality}
		\sum^{d}_{i=1} \Big(\psi_{1}\Big(\frac{\nu\!-\!d\!-\!i}{2}\Big) - \frac{2}{\nu\!-\!d\!-\!i} \Big) > 0.
		\end{equation}
		A sufficient condition for the inequality to hold is for the difference within the summation to remain positive.
		That is, for the inequality
		\begin{equation}\label{inequalityTriGamma}
		\psi_{1}(a_{i}) - \frac{1}{a_{i}} > 0
		\end{equation}
		to hold true for all $a_{i} = \frac{1}{2}(\nu\!-\!d\!-\!i)$, where $i = 1,\ldots,d$.	
		By substitution of recurrence relation $\psi_{1}(a_{i}\!+\!1) = \psi_{1}(a_{i}) - a^{-2}_{i}$ into~\cite[Lemma 10]{Yang2014},
		\begin{equation}
		\psi_{1}(a_{i}) > \frac{1}{a^{2}_{i}} + \frac{2a_{i} + 1}{2a^2_{i} + 2a_{i} + \frac{2}{3}} > \frac{1}{a_{i}}, \quad \forall a_{i} > 0.
		\end{equation}
		Hence, inequality~(\ref{inequalityTriGamma}) holds for all $\nu > 2d$.
	\end{proof}
\end{lemma}


\begin{corollary}\label{cor:TrigammaInequality}
	Let $V \in \mathbb{S}^{d}_{++}$ and $\nu > 2d$ where $d \in \mathbb{N}$.
	Then,
	\begin{multline}\label{eq:TrigammaInequalityII}
	\sum\limits_{i=1}^{d} \psi_{1}\Big(\frac{\nu\!-\!d\!-\!i}{2}\Big) > \frac{2}{\nu\!-\!d\!-\!1} \\ \times\text{vec}(V^{-1})^{T}(V \otimes V)\text{vec}(V^{-1}).
	\end{multline}
	\begin{proof}
		Substitute Lemma~\ref{lem:VecValue} into Lemma~\ref{lem:TrigammaInequality}.
	\end{proof}
\end{corollary}


\begin{lemma}\label{lem:MultivariateGammaDerivative}
	Let the scalar $\nu > 2s + 2d$, where $s \in \mathbb{R}$ and $d \in \mathbb{N}$.
	Then,
	\begin{multline}\label{eq:A-MultivariateGammaDerivative}
	\frac{\partial}{\partial s}\Gamma_{d}\Big(\frac{\nu\!-\!2s\!-\!d\!-\!1}{2}\Big) = -\Gamma_{d}\Big(\frac{\nu\!-\!2s\!-\!d\!-\!1}{2}\Big) \\ \times\sum^{d}_{i=1} \psi_{0}\Big(\frac{\nu\!-\!2s\!-\!d\!-\!i}{2}\Big).
	\end{multline}	
	\begin{proof}
		In accordance with logarithmic differentiation,
		\begin{multline}
		\frac{\partial}{\partial s}\Gamma_{d}\Big(\frac{\nu\!-\!2s\!-\!d\!-\!1}{2}\Big) = \Gamma_{d}\Big(\frac{\nu\!-\!2s\!-\!d\!-\!1}{2}\Big) \\ \times\frac{\partial}{\partial s}\ln\!\Big(\Gamma_{d}\Big(\frac{\nu\!-\!2s\!-\!d\!-\!1}{2}\Big)\Big), \label{eq:GammaPartialOne}
		\end{multline}
		where, via~\cite[Theorem 1.4.1]{Gupta2000},
		\begin{subequations}			
			\begin{align}
			\begin{split}
			\!\!\frac{\partial}{\partial s}\!\ln\!\Big(\Gamma_{d}\Big(\frac{\nu\!-\!2s\!-\!d\!-\!1}{2}\Big)\Big) \!&= \! \frac{\partial}{\partial s} \frac{d(d\!-\!1)}{4}\ln(\pi) + \\ \sum^{d}_{i=1} \frac{\partial}{\partial s}&\ln\!\Big(\Gamma \Big(\frac{\nu\!-\!2s\!-\!d\!-\!i}{2}\Big)\Big),
			\end{split} \\
			&=\! -\sum^{d}_{i=1} \!\psi_{0}\Big(\frac{\nu\!-\!2s\!-\!d\!-\!i}{2}\Big). \label{eq:logGamDeriv}
			\end{align}
		\end{subequations}
	Substitution of~(\ref{eq:logGamDeriv}) into~(\ref{eq:GammaPartialOne}) yields~(\ref{eq:A-MultivariateGammaDerivative}).
	\end{proof}
\end{lemma}


\begin{lemma}\label{lem:ExpectationDS}
	Let $s \in \mathbb{R}$, and $V \in \mathbb{S}^{d}_{++}$ where $d \in \mathbb{N}$.
	Then,
	\begin{equation}
	\frac{\partial}{\partial s}\frac{|M_{\mathbf{x}}\bar{V}M^{T}_{\mathbf{x}}|^{s}}{2^{ds}} = \ln\bigg(\frac{|M_{\mathbf{x}}\bar{V}M^{T}_{\mathbf{x}}|}{2^{d}}\bigg) \frac{|M_{\mathbf{x}}\bar{V}M^{T}_{\mathbf{x}}|^{s}}{2^{ds}}
	\end{equation}
	\begin{proof}
		Given $a^{s} = \exp(s\ln(a))$,
		\begin{subequations}
			\begin{align}
			\frac{\partial}{\partial s}\frac{|M_{\mathbf{x}}\bar{V}M^{T}_{\mathbf{x}}|^{s}}{2^{ds}} &= \frac{\partial}{\partial s} \exp\bigg(\!s \ln\bigg(\frac{|M_{\mathbf{x}}\bar{V}M^{T}_{\mathbf{x}}|}{2^{d}}\bigg)\!\bigg), \\
			&= \ln\bigg(\frac{|M_{\mathbf{x}}\bar{V}M^{T}_{\mathbf{x}}|}{2^{d}}\bigg)\frac{|M_{\mathbf{x}}\bar{V}M^{T}_{\mathbf{x}}|^{s}}{2^{ds}}.
			\end{align}
		\end{subequations}
	\end{proof}
\end{lemma}


\begin{lemma}\label{lem:VolumeResults}
	Let the state transition degrees of freedom $v_{k+1}$ be defined by~(\ref{eq:nuSetting1}).
	Then,
	\begin{equation}
	\begin{split}\label{eq:VolPreserve3}
	&\textup{Vol}\big(\mathbb{E}[X_{k+1}|\mathbf{Z}^{k}]\big) = \\&\textup{Vol}\bigg(\!\int \!M_{\mathbf{x}_{k}}\mathbb{E}[X_{k}|\mathbf{Z}^{k}]H  M_{\mathbf{x}_{k}}^{T} \mathcal{N}(\mathbf{x}_{k}|\mathbf{m}_{k|k},P_{k|k}) d\mathbf{x}_{k} \bigg),
	\end{split}
	\end{equation}
	where $H = (I_{d}+Q_{k+1}V_{k|k})^{-1}|I_{d}+Q_{k+1}V_{k|k}|^{\frac{1}{d}} \in \mathbb{SL}(d,\mathbb{R})$.
	\begin{proof}		
		Let $\bar{\mathbb{E}}[X_{k+1}|\mathbf{Z}^{k}])$ denote the expected value of~(\ref{eq:PredictionInterim}).
		In accordance with~\cite[Theorem 3.4.3]{Gupta2000}
		\begin{equation}\label{eq:ExpFirstEquation}
		\bar{\mathbb{E}}[X_{k+1}|\mathbf{Z}^{k}] \!=\!\! \int \!\frac{M_{\mathbf{x}_{k}}\bar{V}_{k+1} M^{T}_{\mathbf{x}_{k}}}{v_{k+1}\!-\!2d\!-\!2} \mathcal{N}(\mathbf{x}_{k}|\mathbf{m}_{k|k},P_{k|k}) d\mathbf{x}_{k}. 
		\end{equation}
		However, from~(\ref{eq:VInterm}) and~(\ref{eq:nuSetting1}),
		\begin{subequations}
		\begin{align}
		\frac{\bar{V}_{k+1}}{v_{k+1}\!-\!2d\!-\!2} &= \frac{V_{k|k}(I_{d}+Q_{k+1}V_{k|k})^{-1}}{(v_{k|k}\!-\!2d\!-\!2)|I_{d}+Q_{k+1}V_{k|k}|^{-\frac{1}{d}}}, \\
		&= \mathbb{E}[X_{k}|\mathbf{Z}^{k}]\frac{(I_{d}+Q_{k+1}V_{k|k})^{-1}}{|I_{d}+Q_{k+1}V_{k|k}|^{-\frac{1}{d}}}. \label{eq:Vepxrssion}
		\end{align}
		\end{subequations}
		Substituting~(\ref{eq:Vepxrssion}) into~(\ref{eq:ExpFirstEquation}) and defining the special linear group matrix $H = (I_{d}+Q_{k+1}V_{k|k})^{-1}|I_{d}+Q_{k+1}V_{k|k}|^{\frac{1}{d}}$, the expected value becomes:
		\begin{equation}
		\begin{split}
		\bar{\mathbb{E}}[X_{k+1}|\mathbf{Z}^{k}] &=\!\! \int \!M_{\mathbf{x}_{k}}\mathbb{E}[X_{k}|\mathbf{Z}^{k}]H M^{T}_{\mathbf{x}_{k}}\\ &\qquad\qquad\qquad \times \mathcal{N}(\mathbf{x}_{k}|\mathbf{m}_{k|k},P_{k|k}) d\mathbf{x}_{k}. \label{eq:FinalExpectation}
		\end{split}
		\end{equation}
		Finally, from Theorem~\ref{thm:NuPred},	the following equality holds true:
		\begin{equation}\label{eq:ExpectationsEqual}
		\text{Vol}(\mathbb{E}[X_{k+1}|\mathbf{Z}^{k}]) = \text{Vol}\big(\bar{\mathbb{E}}[X_{k+1}|\mathbf{Z}^{k}]\big).
		\end{equation}
		Substitution of~(\ref{eq:FinalExpectation}) into~(\ref{eq:ExpectationsEqual}) yields~(\ref{eq:VolPreserve3}).
	\end{proof}
\end{lemma}


\bibliographystyle{IEEEtran}

\end{document}